\documentclass[11pt]{article}
\usepackage{graphicx,amssymb,bm}
\title{\bf A Feynman integral in Lifshitz-point and Lorentz-violating theories in
$\mathbb R^D\oplus\mathbb R^m$}
\author{\sc R.B. Paris$^a$ and M.A. Shpot$^b$\\
\\
${}^a\!$ {\em Division of Computing and Mathematics, Abertay University,}\\
{\em Dundee DD1 1HG, UK}\\
{\em E-Mail: r.paris@abertay.ac.uk}\\
${}^b\!$ {\em Institute for Condensed Matter Physics, 79011 Lviv, Ukraine}\\
{\em E-Mail: shpot.mykola@gmail.com}}
\begin{document}
\newcommand{\bee}{\begin{equation}}
\newcommand{\ee}{\end{equation}}
\def\f#1#2{\mbox{${\textstyle \frac{#1}{#2}}$}}
\def\dfrac#1#2{\displaystyle{\frac{#1}{#2}}}
\newcommand{\fr}{\frac{1}{2}}
\newcommand{\fs}{\f{1}{2}}
\newcommand{\g}{\Gamma}
\newcommand{\br}{\biggr}
\newcommand{\bl}{\biggl}
\newcommand{\ra}{\rightarrow}
\renewcommand{\topfraction}{0.9}
\renewcommand{\bottomfraction}{0.9}
\renewcommand{\textfraction}{0.05}
\newcommand{\mcol}{\multicolumn}
\date{}
\maketitle
\pagestyle{myheadings}
\markboth{\hfill {\it R.B. Paris and M.A. Shpot} \hfill}
{\hfill {\it A Feynman integral} \hfill}
\begin{abstract}
We evaluate a one-loop, two-point, massless Feynman integral $I_{D,m}(p,q)$
relevant for perturbative field theoretic calculations in strongly anisotropic
$d=D+m$ dimensional spaces given by the direct sum $\mathbb R^D\oplus\mathbb R^m$.
Our results are valid in the whole convergence region of the integral for
generic (non-integer) co-dimensions $D$ and $m$.
We obtain series expansions of $I_{D,m}(p,q)$ in terms of powers of the variable
$X:=4p^2/q^4$, where $p=|\bm p|$, $q=|\bm q|$, $\bm p\in\mathbb R^D$, $\bm
q\in\mathbb R^m$,
and in terms of generalised hypergeometric functions $_3F_2(-X)$, when $X<1$.
These are subsequently analytically continued to the complementary region $X\ge 1$.
The asymptotic expansion in inverse powers of $X^{1/2}$ is derived.
The correctness of the results is supported by agreement with previously known
special cases and extensive numerical calculations.

\vspace{0.4cm}

\vspace{0.3cm}

\end{abstract}

\vspace{0.3cm}

\noindent $\,$\hrulefill $\,$

\vspace{0.2cm}

\begin{center}
{\bf 1. \  Introduction}
\end{center}
\setcounter{section}{1}
\setcounter{equation}{0}
\renewcommand{\theequation}{\arabic{section}.\arabic{equation}}
We consider the calculation of the Feynman integral
\bee\label{e11}
I_{D,m}(p,q)=\int\frac{d^mq'}{(2\pi)^m}\!\int\frac{d^D p'}{(2\pi)^D}\,
\frac{1}{p'^2+q'^4}\,
\frac{1}{|{\bm p'}+{\bm p}|^2+|{\bm q'}+{\bm q}|^4}
\ee
over $d=D+m$ dimensional space, where the vectors
$({\bm p'},{\bm q'})$, $({\bm p},{\bm q}) \in \mathbb R^D\oplus\mathbb R^m$.
This integral is of importance in the field theoretical treatment of
$m$-axial Lifshitz points \cite{HLS75} in strongly anisotropic $d$-dimensional systems, where
it arises both in the context
of the dimensional expansions in the neighbourhood of the
upper critical dimension of its convergence domain (see \cite{HLS75,SG78,MC99,DS00,SD01})
and of the large-$n$ expansion (see \cite{SPD05,SDP08,SP12}),
$n$ being the number of components of the order-parameter field.
The integral (\ref{e11}) also appears in Lorentz-violating quantum field theories
\cite{Ans08} (see also \cite{AnsHal07,Vis09,SP12}),
where the models and mathematics are often very similar to those used in
Lifshitz-point theory in statistical physics.
The parallels with Lifshitz points have given rise to the currently very popular
directions in quantum gravity; for a very recent review see \cite{Wang17}.

The dimensional space of the integral (\ref{e11}) consists of two complementary Euclidean subspaces $\mathbb R^D$ and $\mathbb R^m$.
Their dimensions, $D\ge 0$ and $m\ge 0$ are considered to be continuous real parameters;
by definition, neither of them can exceed $d$.
It is accepted that each of the subspaces $\mathbb R^D$ and $\mathbb R^m$
has full rotational symmetry.
This implies that the function $I_{D,m}(p,q)$ in (\ref{e11}) depends on the absolute values $p$ and $q$
of the vectors ${\bm p}\in \mathbb R^D$ and ${\bm q}\in \mathbb R^m$.
The situation with a broken rotational symmetry in $\mathbb R^m$ when $m\ge2$
has been considered in \cite{DSZ03}.\footnote{Another calculation that allowed the
breaking of the $O(m)$ symmetry can be found  in \cite{IHB90} for a very specific case with $m=d$.
}
The ${\cal D}:=\{D,m\}$ dimensional integrals are taken over infinite Euclidean spaces
$\mathbb R^{\cal D}$, which means
\bee\label{E12}
\int d^{\cal D}k:=\int_{\mathbb R^{\cal D}} d^{\cal D}k=
\int_{-\infty}^\infty dk_1\cdots\int_{-\infty}^\infty dk_{\cal D},
\ee
where the vector $\bm k$ with components $k_1,\ldots,k_{\cal D}$
stands for either of the integration variables $\bm p'$ or $\bm q'$ in (\ref{e11}).
Since the components of both integration variables run from $-\infty$ to $+\infty$,
any finite additive shift in $\bm p'$ or $\bm q'$ leaves the integral (\ref{e11})
unchanged.

The integral (\ref{e11}) converges inside a domain of the $(m,d)$ plane bounded by
the lines $m=0$, $d=m$ (that is $D=0$) and the lower and upper boundaries
$d_\ell\equiv d_\ell(m)=2+\fs m$ ($2\le d_\ell\le 4$)
and $d_u\equiv d_u(m)=4+\fs m$ ($4\le d_u\le 8$), respectively; see Fig.~1.
The function $I_{D,m}(p,q)$ can be extended outside this convergence domain by analytic continuation. Deviations from the upper and lower boundaries are defined by
\bee\label{e110}
\epsilon:=d-2-\fs m=D-2+\fs m,\qquad \epsilon':=4+\fs m-d=4-D-\fs m,
\ee
where $\epsilon'=2-\epsilon$. Both $\epsilon$ and $\epsilon'$ lie in the interval $(0,2)$.
As one approaches the lower boundary ($\epsilon\to 0+$) the integral (\ref{e11}) diverges for small (finite) integration variables, whereas as one approaches the upper boundary
($\epsilon'\to 0+$) the integral diverges for large integration variables. In the neighbourhood of these boundaries, these divergences manifest themselves as simple poles with behaviour proportional to $1/\epsilon$ and $1/\epsilon'$, respectively; see Section 2.
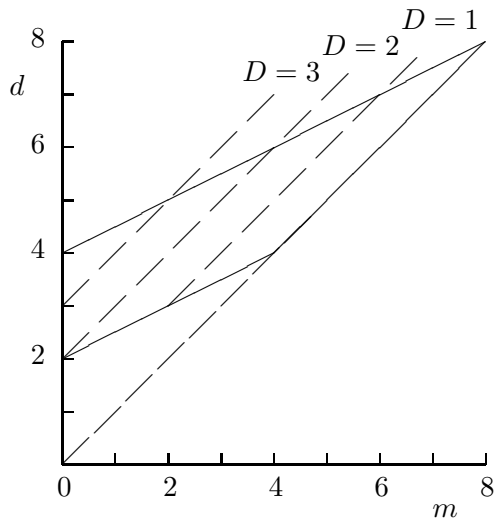
\begin{figure}[t]
\centering
\begin{picture}(200,200)(-25,-25)
\put(0,0){\line(1,0){160}}
\put(0,0){\line(0,1){160}}
\multiput(0,0)(12,12){8}{\line(1,1){10}}
\put(80,80){\line(1,1){80}}
\put(0,40){\line(2,1){80}}
\put(0,80){\line(2,1){160}}
\multiput(0,40)(14,14){8}{\line(1,1){10}}
\multiput(0,60)(14,14){6}{\line(1,1){10}}
\multiput(40,60)(14,14){7}{\line(1,1){10}}
\put(0,20){\line(1,0){4}}
\put(0,40){\line(1,0){4}}
\put(0,60){\line(1,0){4}}
\put(0,80){\line(1,0){4}}
\put(0,100){\line(1,0){4}}
\put(0,120){\line(1,0){4}}
\put(0,140){\line(1,0){4}}
\put(0,160){\line(1,0){4}}
\put(20,0){\line(0,1){4}}
\put(40,0){\line(0,1){4}}
\put(60,0){\line(0,1){4}}
\put(80,0){\line(0,1){4}}
\put(100,0){\line(0,1){4}}
\put(120,0){\line(0,1){4}}
\put(140,0){\line(0,1){4}}
\put(160,0){\line(0,1){4}}
\put(-2,-12){0}
\put(38,-12){2}
\put(78,-12){4}
\put(118,-12){6}
\put(158,-12){8}
\put(-12,38){2}
\put(-12,78){4}
\put(-12,118){6}
\put(-12,158){8}
\put(-20,140){$d$}
\put(140,-20){$m$}
\put(128,165){$D=1$}
\put(98,155){$D=2$}
\put(68,145){$D=3$}
\end{picture}
\caption{\footnotesize{The domain of convergence in the $(m,d)$ plane bounded by $m=0$, $d=m$ and the lower and upper boundaries $d_\ell=2+m/2$ and $d_u=4+m/2$, respectively.
The lines $d=m+D$, with $D=0, 1, 2, 3$, are shown by dashed lines.}}
\end{figure}

The possession of information on the behaviour of the integral $I_{D,m}(p,q)$
in the vicinity of the boundary lines
of the critical dimensions $d_u(m)=\fs m+4$ and $d_\ell(m)=\fs m+2$ (see Fig. 1)
has allowed the achievement of several important results.
(i) The agreement between the large-$n$ expansion and the
dimensional expansion near $d_u(m)$
for the correlation critical exponents $\eta_{L2}$ and $\eta_{L4}$ has been explicitly shown for generic $m$
at $d_u(m)-d:=\epsilon'\to0$ in \cite{SDP08}.
(ii) A similar consistency check between these exponents at large $n$ and at low
dimensions $d=d_\ell(m)+\epsilon$ with $\epsilon\to0$ (obtained in \cite{GS78}) has been carried out in
\cite[Sec. 6]{SPD05}.
(iii) A consideration of the large-$n$ expansions of $\eta_{L2}$ and $\eta_{L4}$
at $m=1$ near the critical dimensions $d_u(1)=4.5$ and
$d_\ell(1)=2.5$, combined with their knowledge at certain other points of the $(m,d)$ plane,
has suggested schematic plots
for the $O(1/n)$ coefficients of these exponents as functions of the
space dimensionality $d$; see \cite[Fig. 3]{SPD05}. Both of these curves appeared to show non-trivial
non-monotonic behaviour.

On the other hand,
recently the (multi)critical behaviour at Lifshitz points has also been studied
by means of the functional renormalisation group equations \cite{Ber04,EKM12,BZ15,Zap17}%
\footnote{References \cite{Ber04,EKM12} deal with \emph{anisotropic} Lifshitz points,
for which generically $0<m<d$, while \cite{BZ15,Zap17} consider their \emph{isotropic}
analogue \cite{HLS75n,ES77,DS02,Gubser17} with $m=d$.
The essentially more complicated model that allows anisotropy along all $d$ spatial directions
has been studied by means of the large-$n$ expansion in \cite{IHB90}.}.
The advantage of this approach is that it gives \emph{direct} access both to generic
dimensions ($m$, $d$) and numbers of order parameter components $n$.
Thus, in \cite[Fig. 1]{EKM12}, the exponents
$\eta_{L2}$ and $\eta_{L4}$ of the uniaxial ($m=1$) Lifshitz point with $n=3$
have been plotted as continuous functions of the space dimension $d$ for
$2.5\le d\le 4.5$, which corresponds to the whole interval between the lower and upper boundaries
$d_\ell(1)$ and $d_u(1)$.
These graphs reproduced the main qualitative features predicted in the schematic
plots drawn in \cite[Fig. 3]{SPD05} for the $O(1/n)$ coefficients of these exponents
in the large-$n$ expansion.
In fact, it has been anticipated in \cite{SPD05} that the dimensional dependence
of each of the exponents $\eta_{L2}$ and $\eta_{L4}$
will maintain its shape given by Fig. 3 of this reference beyond the assumption
of $n\to\infty$ and $m=1$; this conjecture has been confirmed by the results of \cite{EKM12}.

For general $m$ and $d$ in the (physical) region indicated in Fig.~1,
the correlation critical exponents $\eta_{L2}$ and $\eta_{L4}$
have been found in \cite{SPD05} to order $O(1/n)$ of the large-$n$ expansion.
These results have been expressed in the form of integrals whose integrands involve
the function $I_{D,m}(p,q)$ given by (\ref{e11}) in their denominators
(see \cite[(36), (38)]{SPD05}).
Thus, knowledge of the integral $I_{D,m}(p,q)$ at
any $D$ and $m$ in its convergence region 
for arbitrary $p$ and $q$ from zero to infinity
would allow the construction of detailed plots of the dimensional dependence of the large-$n$
coefficients of $\eta_{L2}$ and $\eta_{L4}$ starting from their integral representations
\cite[(36), (38)]{SPD05}.
This would lead also to the prediction of the dependence on $d$ of the important
\emph{non-trivial} anisotropy index $\theta=(2-\eta_{L2})/(4-\eta_{L4})$;
see \cite[(37)]{SPD05}.%
\footnote{
The inverse of this value, $1/\theta$, corresponds to the non-trivial dynamical critical exponent
of the \emph{renormalised} $z=2$ scalar Lorentz violating theory,
$z$ being the effective-Hamiltonian parameter appearing as a power of $q^2$ in $(q^2)^z$;
cf. \cite{Ans08}, \cite[p.78]{SP12}.
}
Such a calculation of $I_{D,m}(p,q)$ would also provide the first step in the study of
the analogous dimensional dependence of the \emph{thermodynamic} critical exponents
derived at order $1/n$ in \cite{SP12} (however, another, more involved integral
of the type given in (\ref{e11}) would be needed in this case).

Thus, one of the main motivations of the present work has been to
provide new possibilities for
further investigation of the dimensional dependence of critical exponents
of $m$-axial Lifshitz points in $d$ spatial dimensions
and for further comparison of the results of the $1/n$ expansion in \cite{SPD05,SP12}
with those of \cite{EKM12}.
Another aspect of our motivation was to calculate the integral $I_{D,m}(p,q)$
for \emph{all} dimensions $D$ and $m$ in its convergence domain.
This has never been done before.
Having expressions that cover the whole convergence domain of the
integral in (\ref{e11}) would then allow to test the results at all four
boundary lines (see Fig.~1), as well as for all special cases that have been previously calculated
by different means (see Section 2).
The ability to perform all possible checks provides good confidence in the obtained results.
Such an approach -- of working with functions depending on generic parameters
allowed by the theory -- made it possible in \cite{DS00,SD01,SPD05,SP12}
to consider all required limiting and special cases. In turn,
this has then led to the possibility of producing, with confidence,
the correct results and sorting out (numerous) erroneous results and approaches
in the Lifshitz-point theory;
see \cite{DS00,SD01,SP12}, as well as \cite{DS01,DS03}, and the references therein.

\vspace{0.6cm}

\begin{center}
{\bf 2. \  Some general forms and special evaluations}
\end{center}
\setcounter{section}{2}
\setcounter{equation}{0}
\renewcommand{\theequation}{\arabic{section}.\arabic{equation}}
The integral $I_{D,m}(p,q)$ is a generalised homogeneous function, which implies that a power of each argument,
$p^{-2+\epsilon}$ or $q^{-4+2\epsilon}$ can be scaled out. This results in the following two equivalent scaled forms
\bee\label{e12}
I_{D,m}(p,q)=p^{-2+\epsilon} I_{D,m}\Big(1,\frac{q}{\sqrt{p}}\Big):=
p^{-2+\epsilon}f_1\Big(\frac{q}{\sqrt{p}}\Big)
\ee
and
\bee\label{f12}
I_{D,m}(p,q)=q^{-4+2\epsilon} I_{D,m}\Big(\frac{p}{q^2},1\Big):=
q^{-4+2\epsilon} f_2\Big(\frac{p}{q^2}\Big).
\ee
By comparing (\ref{e12}) and (\ref{f12}) it is easy to see that the scaling functions
$f_1$ and $f_2$ are related via the functional equation
\[
f_2(u)=u^{-2+\epsilon}f_1\Big(\frac1{\sqrt u}\Big).
\]

Further, it follows that the values of $I_{D,m}(p,q)$ at zero arguments are given by
\bee\label{e13}
I_{D,m}(p,0)=p^{-2+\epsilon} I_{D,m}(1,0)\quad\mbox{and}\quad I_{D,m}(0,q)=q^{-4+2\epsilon} I_{D,m}(0,1),
\ee
where (obtained by different means in \cite[(B.13)]{SD01} and \cite[(B.8)]{SPD05})
\bee\label{e14a}
I_{D,m}(1,0)=\frac{2^{\epsilon}\,\g(\fs\epsilon')}
{(16\pi)^{(d-1)/2}\g\left(\frac{3-D+\epsilon}2\right)}\,
\frac{\g^2(\fs\epsilon)}{\g(\epsilon)}\,,
\ee
and (cf. \cite[(B.23)]{SPD05})
\bee\label{e14b}
I_{D,m}(0,1)=\frac8{(16\pi)^{(d-1)/2}}\,\frac\pi{\sin\pi\epsilon}\,\frac1{2{-}D}
\bl[\frac1{\Gamma\left(\frac{1-D}2+\epsilon\right)}
-\frac{\sqrt\pi}{\Gamma\left(-\frac{1}2+\epsilon\right)\Gamma\left(\frac{3-D}2\right)}\br].
\ee
Here we have written the last two expressions in a form that easily allows to see
the presence of the dimensional poles $\sim1/\epsilon$ and $\sim1/\epsilon'$ near the
lower and upper boundaries $d_\ell$ and $d_u$.
In (\ref{e14a}) these singularities are given by the poles of the
gamma functions depending on $\epsilon$ and $\epsilon'$, respectively.
In (\ref{e14b}) both poles are contained in the factor
$\pi/\sin\pi\epsilon=-\pi/\sin\pi\epsilon'$.
The singularity resulting from the factor $1/(2-D)$ is apparent, since the content of
the square brackets in (\ref{e14b}) vanishes at $D=2$.
The expressions in (\ref{e13})--(\ref{e14b}) enable the determination of the asymptotic behaviour of $I_{D,m}(p,q)$ as either $p\to\infty$ or $q\to\infty$ when the remaining variable is finite.
In the following, we shall check our calculations against (\ref{e14a}) and (\ref{e14b}).

There is another role played by the special values $I_{D,m}(p,0)$ and $I_{D,m}(0,q)$.
They are directly related to the marginal cases of the integral (\ref{e11})
when $m=0$ and $D=0$.
In these cases the corresponding subspaces $\mathbb R^m$ and $\mathbb R^D$ disappear,
along with all vectors defined therein, while their complements become $\mathbb R^d$.
Thus, at $m=0$  the integral $I_{D,m}(p,q)$ simplifies to
\bee\label{ed0}
I_{d,0}(p,0)=\int\frac{d^d p'}{(2\pi)^d}\frac{1}{p'^2({\bm p'}+{\bm p})^2}=
\frac{p^{d-4}}{(4\pi)^{d/2}}\,\frac{\g^2(\fs d-1)\g(2-\fs d)}{\g(d-2)}.
\ee
This is a textbook example that can be found, for example, in \cite[p. 160]{ZJ89}.
On the other hand, when $D=0$ we have
\bee\label{e0d}
I_{0,d}(0,q)=\int\frac{d^d q'}{(2\pi)^d}\frac{1}{q'^4({\bm q'}+{\bm q})^4}=
\frac{q^{d-8}}{(4\pi)^{d/2}}\,\frac{\g^2(\fs d-2)\g(4-\fs d)}{\g(d-4)},
\ee
which coincides with \cite[(A.1)]{DS02}.
It is then straightforward to see that with $m=0$, $D=d$, $\epsilon=d-2$ and $\epsilon'=4-d$
(recall (\ref{e110})), (\ref{e14a}) reduces to the coefficient of $p^{d-4}$ in (\ref{ed0}).
Similarly, when $D=0$, $m=d$ and $\epsilon=d/2-2$,
(\ref{e14b}) agrees with (\ref{e0d}).

For several pairs of $D$ and $m$, which represent certain \emph{points} in the
$(m,d)$ plane in Fig.~1, the function $I_{D,m}(p,q)$ has been calculated explicitly
some time ago. We define the quantity
\bee\label{e120}
X:=\frac{4p^2}{q^4}~,
\ee
which will be employed throughout the sequel. Then for $X>0$, we have,
for $D=1,2,3$ and different $m$,
\bee\label{c17}
I_{1,4}(p,q)=\frac{q^{-2}}{32\pi^2} \bl\{\log \frac{X+4}{X+1}+\frac{2}{\sqrt{X}} \,\arctan\,\frac{X^{3/2}}{3X+4}\br\}
\qquad\mbox{\cite[(68)]{SPD05}, \cite[(3.9)]{SP16}};
\ee
\bee\label{e17}
I_{2,1}(p,q)=p^{-3/2}\,\frac{w^{1/4}}{4\sqrt{2}}\;{}_2F_1(\f{1}{4},\f{1}{4};1;w)
\quad\mbox{with}\quad w=\frac{X^3}{(X+4)^2(X+1)}
\quad\mbox{\cite[(75)]{SPD05}}
\ee
derived in \cite{SPD05} using a calculation carried out in the complex plane by S. Rutkevich%
\footnote{Private communication to MAS.};
\bee\label{g17}
I_{3,1}(p,q)=\frac{q^{-1}}{4\pi\sqrt 2}\frac1{\sqrt{\sqrt{X+1}+1}}
\qquad\mbox{\cite[(65)]{SPD05}, \cite[(A.3)]{Ans08}}.
\ee

Explicit expressions for $I_{D,m}(p,q)$ in terms of Gauss hypergeometric functions
along two \emph{lines} in the $(m,d)$ plane appeared quite recently in \cite{SP16}.
These are:
\bee\label{e15}
I_{1,m}(p,q)=q^{-4+2\epsilon}\,\frac{\g(1-\epsilon)}{\pi\epsilon(16\pi)^\epsilon}\,
\frac1{\sqrt X}\,\Im \bl\{(1-i\sqrt X)^{-1+\epsilon}
{}_2F_1\bl(1,1-\epsilon;1+\epsilon;\frac{-1}{1-i\sqrt X}\br)\br\}
\ee
when $D=1$ and $m\in(2,6)$, where from (\ref{e110}) $\epsilon=\fs m-1 \in(0,2)$;
and
\bee\label{e16}
I_{3,m}(p,q)=q^{-4+2\epsilon}\,\frac{8\,\Gamma(2-\epsilon)}{(16\pi)^\epsilon}
(1+X)^{-1+\frac\epsilon2}\,
{}_2F_1\Big(1-\frac{\epsilon}{2},\frac{\epsilon}{2}; \frac32;\frac{X}{1+X}\Big)
\ee
when $D=3$ and $m \in (0,2)$, where $\epsilon=1+\fs m \in[1,2)$; see Fig.~1.
Equations (\ref{c17}) and (\ref{g17}) are special cases of
(\ref{e15}) and (\ref{e16}), respectively.
Further specialisations of (\ref{e15}) for $m=3$ and 5 are given in \cite{SP16}.
In addition,  the Taylor expansion of
$I_{D,1}(p,q)$ in powers of $X$ on the \emph{vertical line } $m=1$ has been obtained in \cite[Sec. 5.3]{RDS11}.

\vspace{0.6cm}

\begin{center}
{\bf 3. \  Evaluation of $I_{D,m}(p,q)$ in the general case}
\end{center}
\setcounter{section}{3}
\setcounter{equation}{0}
\renewcommand{\theequation}{\arabic{section}.\arabic{equation}}
The inner integral over ${\bm p'}$ in (\ref{e11}) has the form
\bee\label{JD}
J_D(p\,; m_1, m_2)=\int\frac{d^Dp'}{(2\pi)^D}\,
\frac{1}{p'^2+m_1^2}\,\frac{1}{|{\bm p'}+{\bm p}|^2+m_2^2}.
\ee
This is the standard $D$-dimensional
one-loop two-point Feynman integral with external momentum ${\bm p}$
and arbitrary ``masses" $m_1$ and $m_2$.
In \cite{Sh07}, this integral has been evaluated as%
\footnote{
In the special case $m_1=0$ and $m_2:=\kappa$, (\ref{e21}) reduces to
\[
J_D(p\,;0,\kappa)=J_D(p;\kappa,0)=-\frac{2\,\gamma_D}{2-D}\kappa^{D-4}\,
_2F_1\Big(2-\frac{D}{2},1;\frac{D}{2};-\frac{p^2}{\kappa^2}\Big)
\]
which agrees with \cite[(10)]{BD91} taken at $\alpha=\beta=1$.
The factor $2/(2-D)$ is missing in the misprinted formula \cite[(2.7)]{SP16}.
}
\bee\label{e21}
J_D(p\,;m_1, m_2)=\gamma_D\Big(\frac{m_1+m_2}2\Big)^{D-4}F_1\bl(\frac{4{-}D}2;1,\frac{3{-}D}2;
\frac32;\frac{-p^2}{(m_1+m_2)^2}, \frac{(m_2-m_1)^2}{(m_1+m_2)^2}\br),
\ee
where
$$\gamma_D:=(4\pi)^{-D/2} \g(2-\fs D).$$
The function $F_1$  denotes the Appell hypergeometric function
defined by the double series expansion \cite[p.~224]{EMOT1}, \cite[p.~53]{SriMan}, \cite[p.~22]{SriKar}
\[F_1(\alpha;\beta,\beta';\gamma;x,y)=\sum_{k\geq0}\sum_{n\geq0} \frac{(\alpha)_{k+n}(\beta)_k(\beta')_n}{(\gamma)_{k+n}}\,\frac{x^k}{k!}\,\frac{y^n}{n!}\qquad (|x|<1, |y|<1),\]
where $(a)_n=\g(a+n)/\g(a)=a(a+1)\ldots (a+n-1)$, $(a)_0=(0)_0:=1$,
is the Pochhammer symbol, or rising factorial.

The integral $J_D(p;m_1,m_2)$ can be expressed alternatively in terms of a linear
combination of two Gauss hypergeometric functions
\cite[(A.7)]{BDS96}, \cite[(7)]{Sh07},
the Appell function $F_4$ \cite[(17)]{Sh07},
Horn function $H_3$ (see \cite[(p. 20)]{Sh07} and \cite[(33)]{SriMan}),
as well as different versions of the Appell function $F_1$:
\cite[(28)]{Sh07}, \cite[(7)]{KT12}, see also \cite[Sec. 2.1]{SP16}.
Calculations of different versions of the Feynman integral $J_D(p;m_1,m_2)$ in (\ref{JD})
using different approaches are continuing to attract attention;
see, for example, \cite{BPR16,Kol16}.
For our present purposes of further integration over $\bm q'$ in (\ref{e11}), we shall use
the expression (\ref{e21}), which leads to most straightforward calculations.

In (\ref{e11}), the ``masses" $m_1$ and $m_2$ 
correspond to $q'^2$ and $({\bm q'}+{\bm q})^2$, respectively.
They are related to the integration variable of the outer integral $\bm q'$.
As mentioned above (see (\ref{E12})), this variable can be shifted at will.
To make the outer integral in (\ref{e11}) more symmetric we can
change the integration variable via ${\bm q}'={\bm y}-{\bm q}/2$.
This implies that
\bee\label{MM}
m_1=({\bm y}-{\bm q}/2)^2\quad\mbox{and}\quad
m_2=({\bm y}+{\bm q}/2)^2
\ee
have to be used in (\ref{e21}). Thus,
the combinations of $m_1$ and $m_2$ appearing in the arguments of the Appell function $F_1$
in (\ref{e21}) are
\[m_2-m_1=2({\bm y}\cdot{\bm q})\quad\mbox{and}\quad
 m_1+m_2=2(y^2+\f{1}{4}q^2).\]

The convergence of the series expansion of $F_1$ in (\ref{e21}) then requires $p^2/(4(y^2+q^2/4)^2)<1$; for $y^2\geq0$, this corresponds to $X<1$
(recall (\ref{e120})).
The restriction $X<1$ could be apparently avoided if we used, as an alternative,
the function $J_D(p\,;m_1,m_2)$ in the form \cite[(28)]{Sh07}.
This is obtained from (\ref{e21}) by a linear transformation of the
Appell function $F_1$, which provides an analytical continuation of
this function necessary for the validity of the results for any $X>0$.
In fact, we find it possible to carry out such integration
in closed form for arbitrary $D$ and $m$ in the convergence domain
of (\ref{e11}).

Thus, our further task is to calculate
\bee\label{r24}
I_{D,m}(p,q)=\int\frac{d^my}{(2\pi)^m}\,J_D(p\,;m_1,m_2)
\ee
with $J_D(p\,;m_1,m_2)$ given in (\ref{e21}) and $m_1$, $m_2$ in (\ref{MM}).
Insertion of (\ref{e21}), using the above double-series definition
of the Appell function $F_1$, into
(\ref{r24}) yields, with the scalar product ${\bm y}\cdot{\bm q}=yq\cos \theta$,
\[
I_{D,m}(p,q)=\frac{\gamma_D}{(2\pi)^m}\int\frac{d^my}{(y^2+\f{1}{4}q^2)^{4-D}}
\sum_{n,k\geq0}(-)^k{\cal A}_{nk} \bl(\frac{p}{2(y^2+\f{1}{4}q^2)}\br)^{2k}
\bl(\frac{yq\cos \theta}{y^2+\f{1}{4}q^2}\br)^{2n}.
\]
For convenience in presentation we have defined here the coefficients ${\cal A}_{nk}$ by
\bee\label{Ank}
{\cal A}_{nk}:=\frac{(\frac{4{-}D}2)_{k+n}(\frac{3{-}D}2)_n}{(\f{3}{2})_{k+n}\,n!}~.
\ee
Then, for the above $m$-dimensional integral%
\footnote{
The volume element is $dV=r^{m-1}\sin^{m-2}\theta_1 \sin^{m-3}\theta_2 \ldots \sin \theta_{m-2}dr d\theta_1\ldots d\theta_{m-1}$ and we obtain (cf. \cite[3.3.1]{PBM1})
\[\int \frac{d^mr}{(2\pi)^m}\,f(r,\theta_1)=\int_0^\infty\!\! \!r^{m\!-\!1}dr \int_0^\pi\!\! \sin^{m\!-\!2}\theta_1 f(r,\theta_1)\,d\theta_1\cdot
\int_0^\pi\!\! \sin^{m\!-\!3}\theta_2 d\theta_2 \ldots \int_0^\pi\!\! \sin \theta_{m\!-\!2} d\theta_{m\!-\!2} \int_0^{2\pi}\!\!d\theta_{m-1}\]
\[=K_m\int_0^\infty\!\! r^{m-1} dr \int_0^\pi \!\frac{\sin^{m\!-\!2}\theta_1}{\Omega_m}\,f(r,\theta_1)\,d\theta_1,\]
where $K_m$ and $\Omega_m$ are the geometric factors introduced in (\ref{e20}).}
we obtain
\[I_{D,m}(p,q)=\gamma_D K_m \sum_{n,k\geq0} {\cal A}_{nk} q^{2n} (-\f{1}{4}p^2)^k
\int_0^\infty
\frac{y^{m+2n-1}}{(y^2+\f{1}{4}q^2)^\omega}\,dy \int_0^\pi \frac{\sin^{m-2}\theta}{\Omega_m} \cos^{2n}\theta\,d\theta,\]
where $\omega:=4-D+2(k+n)$. Here the quantities $K_m$ and $\Omega_m$
are the usual geometric constants given by
\bee\label{e20}
K_m:=\frac{S_m}{(2\pi)^m}=\frac{2\,(4\pi)^{-m/2}}{\g(\fs m)},\qquad
\Omega_m:=\sqrt{\pi}\;\frac{\g(\fs m-\fs)}{\g(\fs m)},
\ee
with $S_m$ denoting the area of a unit sphere in $m$ dimensions, and
$\Omega_m=\int_0^\pi\sin^{m-2}\theta\,d\theta$.

Use of the standard integrals
\[
\int_0^\pi \sin^{m-2}\theta \cos^{2n}\theta\,d\theta=
2\int_0^{\pi/2} \sin^{m-2}\theta \cos^{2n}\theta\,d\theta
=\Omega_m\,\frac{(\fs)_n}{(\frac m2)_n},
\]
\[\int_0^\infty \frac{y^{\nu-1}}{(y^2+\lambda^2)^\mu}\,dy=\lambda^{\nu-2\mu}\,
\frac{\g(\fs\nu)\g(\mu-\fs\nu)}{2\g(\mu)}\qquad (\mu>\fs\nu>0)\]
then shows that
\[
I_{D,m}(p,q)=\frac{\gamma_D K_m}{2\Omega_m}\g\Big(\frac{m{-}1}2\Big)
\Big(\frac{q}{2}\Big)^{m-8+2D}\!\!
\sum_{n,k\geq0}\!\!4^n {\cal A}_{nk}
\frac{\g(n{+}\fs)\g(4{-}D{-}\fs m{+}2k{+}n)}{\g(4-D+2k+2n)}(-X)^k
\]
\bee\label{e22}
=A_{D,m}\,q^{-4+2\epsilon}\, \sum_{n,k\geq0}4^n {\cal A}_{nk}\,\frac{(\fs)_n
(2-\epsilon)_{2k+n}}{(4-D)_{2k+2n}} (-X)^k.
\ee
Here we have recalled the definition of the parameter $\epsilon$ in (\ref{e110}), and,
for the combination of all numerical factors, have introduced the notation $A_{D,m}$ given by
\bee\label{e23}
A_{D,m}:=
\frac{8}{(16\pi)^{\frac{d-1}2}}\,\frac{\g(2-\epsilon)}{\g(\frac{5-D}2)}.
\ee
We  notice that the relations (\ref{e110}) and $d=D+m$ imply that
$(d-1)/2=(3-D)/2+\epsilon$.

For the convenience of presentation, we introduce the further notation given by
\bee\label{e240}
a:=4-D-\fs m=2-\epsilon=\epsilon', \qquad b:=\frac{3-D}2.
\ee
Substitution of the coefficients ${\cal A}_{nk}$ from (\ref{Ank})
into (\ref{e22}), followed by use of the relation
\bee\label{DOU}
(2a)_{2k}=2^{2k}(a)_k(a+\fs)_k,
\ee
then yields the following compact double-sum representation for $I_{D,m}(p,q)$:

\newtheorem{theorem}{Theorem}
\begin{theorem}\label{T1}
$\!\!\!.$\ For any $D$ and $m$ inside the convergence region
of the integral $I_{D,m}(p,q)$ in (\ref{e11}), we have the series expansion
\bee\label{e241}
I_{D,m}(p,q)=A_{D,m}\,q^{-4+2\epsilon}\sum_{n,k\geq0} \frac{(a)_{2k+n} (\fs)_n (b)_n}{(\f{3}{2})_{k+n} (b+1)_{k+n} n!}\,(-X/4)^k \qquad (X<1)
\ee
where $X=4p^2/q^4$, and the coefficient $A_{D,m}$ and the parameters $a$ and $b$
are defined in (\ref{e23}) and (\ref{e240}).
\end{theorem}

If in (\ref{e241}) the summation over $n$ is carried out first, we obtain
\begin{eqnarray}
I_{D,m}(p,q)&=&A_{D,m}\,q^{-4+2\epsilon} \sum_{n,k\geq0}
\frac{(a+2k)_n(a)_{2k}(\fs)_n(b)_n}
{(\f{3}{2}+k)_n(\f{3}{2})_k (b+1+k)_n(b+1)_k n!}\,(-X/4)^k\nonumber\\
&=&A_{D,m}\,q^{-4+2\epsilon} \sum_{k\geq0}\frac{(\fs a)_k(\fs a+\fs)_k}{(\f{3}{2})_k (b+1)_k} (-X)^k \sum_{n\geq0} \frac{(a+2k)_n(\fs)_n(b)_n}{(\f{3}{2}+k)_n (b+1+k)_n n!}.
\nonumber
\end{eqnarray}
Using once again (\ref{DOU}) to transform the coefficient in the sum over $k$ and
expressing the inner sum over $n$ as a ${}_3F_2$ series of unit argument, we obtain
an alternative representation of the series expansion (\ref{e241}):
\begin{theorem}\label{TT}$\!\!\!.$\ For $X<1$ we have the series expansion
\bee\label{e26}
I_{D,m}(p,q)=A_{D,m}\,q^{-4+2\epsilon} \sum_{k\geq0} {\cal C}_k F_k
(-X)^k,
\ee
where
\bee\label{c24}
{\cal C}_k:=\frac{(a)_{2k}4^{-k}}{(\f{3}{2})_k (b+1)_k}~,\qquad F_k:= {}_3F_2\bl(\!\!\begin{array}{c}a+2k,\fs,b\\\f{3}{2}+k,b+1+k\end{array}\!\!;1\br),
\ee
$A_{D,m}$ is given by (\ref{e23}),
and the parameters $a$ and $b$ are defined in (\ref{e240}).
\end{theorem}
The ${}_3F_2(1)$ series in (\ref{c24}) converges since its parametric excess is
$2-a=\epsilon>0$, which corresponds, as before, to $D+\fs m>2$, or $d>d_\ell$.
It has negative integral parameter differences and series of this kind have been studied
recently in \cite{SS15}.
Equation (\ref{e26}) is especially useful for the computation of
$I_{D,m}(p,q)$ when $X\ll 1$ ($q$ large with $p$ fixed).

\vspace{0.3cm}

If, on the other hand, we evaluate the summation in (\ref{e241}) over $k$ first,
using the fact that
\bee\label{e2Poch}
(a)_{2k+n}=(a+n)_{2k} (a)_n=(a+2k)_n (a)_{2k}
\ee
together with (\ref{DOU}), we find
\begin{eqnarray}
I_{D,m}(p,q)&=&A_{D,m}\,q^{-4+2\epsilon}\sum_{n,k\geq0}
\frac{(a+n)_{2k}(a)_n(\fs)_n(b)_n}{(\f{3}{2}+n)_k(\f{3}{2})_n (b+1+n)_k(b+1)_n\,n!}\,(-X/4)^k
\nonumber\\\nonumber
&=&A_{D,m}\,q^{-4+2\epsilon}\sum_{n\geq0}{\cal B}_n
\sum_{k\geq0}\frac{(\frac{a+n}{2})_k (\frac{a+n+1}{2})_k (1)_k}{(\f{3}{2}+n)_k (b+1+n)_k k!}\,(-X)^k\,.
\end{eqnarray}
The coefficients ${\cal B}_n$, which group together all Pochhammer symbols
with index $n$, are given by
\bee\label{b24}
{\cal B}_n:=\frac{(\fs)_n (a)_n (b)_n}{(\f{3}{2})_n (b+1)_n\,n!}\,.
\ee
The sum over $k$ is recognised as a $_3F_2$ function of argument $-X$,
and we arrive at the following theorem:

\begin{theorem}\label{TF}
$\!\!\!.$\ For $X<1$ we have the representation of the integral $I_{D,m}(p,q)$
in the form of a sum of $_3F_2$ functions
\bee\label{e25}
I_{D,m}(p,q)=A_{D,m}\,q^{-4+2\epsilon}\sum_{n\geq0}{\cal B}_n\,
{}_3F_2\bl(\!\!\begin{array}{c}1, \fs(a+n), \fs(a+n+1)\\
\f{3}{2}+n, b+1+n\end{array}\!\!;-X\br),
\ee
where the coefficients ${\cal B}_n$ are defined in (\ref{b24}), with
$A_{D,m}$, $a$ and $b$ given in (\ref{e23}) and (\ref{e240}).
\end{theorem}

It is found numerically that the ${}_3F_2(-X)$ function in (\ref{e25}) possesses a slow monotonic character with increasing $n$ for fixed $X>0$. The convergence of the series (\ref{e25}) is then essentially controlled by the coefficients ${\cal B}_n$, which possess the large-$n$ behaviour
\bee\label{eBn}
{\cal B}_n\sim n^{-1-\epsilon}\qquad (n\to\infty),
\ee
thus requiring $\epsilon>0$ (which corresponds to $d>d_\ell=\fs m+2$) for convergence.

When $X=0$, which corresponds to the special case $p=0$,
all $_3F_2$ functions with zero argument in (\ref{e25}) reduce to $1$.
Then from the expansion (\ref{e25}) it follows that
\[
I_{D,m}(0,1)=A_{D,m} \sum_{n\geq0}{\cal B}_n=
A_{D,m}\,{}_3F_2\bl(\begin{array}{c}a,\fs,b\\\f{3}{2},b+1\end{array}\!\!;1\br).
\]
In the last hypergeometric series of unit argument
two pairs of numerator and denominator parameters differ by unity.
Thus this is summable via \cite[7.4.4.16]{PBM3} (see also \cite[(32)]{SS15} at $n=0$)
and produces
\[
I_{D,m}(0,1)=A_{D,m}\,\frac{b\,\g(1-a)}{2b-1}
\bl\{\frac{\sqrt{\pi}}{\g(\f{3}{2}-a)}-\frac{\g(a)}{\g(1-a+b)}\br\}.
\]
With the value of $A_{D,m}$ from (\ref{e23}) and parameters $a$ and $b$ from
(\ref{e240}), this is easily seen to reduce to (\ref{e14b}) as expected.
A similar check of our calculations at $X=\infty$ can be found in Sec. 4.2.

The expansions in (\ref{e26}) and (\ref{e25}) provide the extension of $I_{D,m}(p,q)$ to non-integer values of $D$ and $m$ in the convergence domain (see Fig. 1).
However, all results of the present section are valid only under the restriction $X<1$.
Since for practical applications, discussed in the Introduction, the function $I_{D,m}(p,q)$
is needed for all $X\in[0,\infty)$, we consider, in the next sections, the possibility
of obtaining $I_{D,m}(p,q)$ for the values $X>1$.

\vspace{0.6cm}

\begin{center}
{\bf 4. \ Continuation of the expansions for $I_{D,m}(p,q)$ into $X>1$ and an asymptotic expansion for $X\to\infty$}
\end{center}
\setcounter{section}{4}
\setcounter{equation}{0}
\renewcommand{\theequation}{\arabic{section}.\arabic{equation}}
The power expansion in (\ref{e26}) and functional expansion in (\ref{e25})
have been derived under the condition that $X<1$,
but analytic continuation can be applied to these sums
to continue the results into the region $X>1$.
We shall carry this out in the next sub-sections and also derive modified versions of these expansions suitable for numerical computation. The modified version of (\ref{e25}) will enable
the determination of the asymptotic expansion of $I_{D,m}(p,q)$ for $X\to\infty$.
\vspace{0.3cm}

\noindent{\bf  4.1.\ \ The expansion (\ref{e25}) and its modified form}
\vspace{0.2cm}

In the $_3F_2$ functions appearing in Theorem \ref{TF}
we make use of the connection formula \cite[(16.8.8)]{DLMF}
\bee\label{ea1}
{}_3F_2\bl(\begin{array}{c}\alpha_1, \alpha_2, \alpha_3\\ \beta_1, \beta_2\end{array}\!\!;-X\br)=\sum_{j=1}^3G_j X^{-\alpha_j} {}_3F_2\bl(\begin{array}{c}\alpha_j, 1-\beta_1+\alpha_j, 1-\beta_2+\alpha_j\\ 1-\alpha_1+\alpha_j,\  *\  , 1-\alpha_3+\alpha_j\end{array}\!\!;-\frac{1}{X}\br)
\ee
provided no two of the $\alpha_j$ differ by an integer, where the asterisk denotes the omission of the term in the denominator parameters corresponding to $\alpha_k=\alpha_j$ ($1\leq k\leq j$) and
\[G_j:=\frac{\prod_{k=1, k\neq j}^3 \g(\alpha_k-\alpha_j)/\g(\alpha_k)}{\prod_{k=1}^2 \g(\beta_k-\alpha_j)/\g(\beta_k)}~.\]
Identification of the parameters appearing in (\ref{e25}) as $\alpha_1=1$, $\alpha_2=\fs(a+n)$, $\alpha_3=\fs(a+n+1)$, $\beta_1=\f{3}{2}+n$ and $\beta_2=b+1+n$ in (\ref{ea1}) then yields the following result.
\begin{theorem}\label{TG}
$\!\!\!.$\ For $X>1$ we have the expansion
\bee\label{e26a}
I_{D,m}(p,q)=A_{D,m}\,q^{-4+2\epsilon}\,\sum_{n\geq0}{\cal B}_n\,\sum_{j=1}^3 G_j(n)f_n^{(j)}(X),
\ee
where
\begin{eqnarray}
f_n^{(1)}(X)&=&X^{-1}{}_3F_2\bl(\begin{array}{c}1, \fs-n, 1-b-n\\ \f{3}{2}-\frac{a+n}{2}, 2-\frac{a+n}{2}\end{array}\!;-X^{-1}\br),\nonumber\\
f_n^{(2)}(X)&=&X^{-(a+n)/2}\,{}_2F_1(\fs(a\!-\!1)\!-\!\fs n,\fs(a\!-\!2b)\!-\!\fs n;\fs;-X^{-1}),\label{e26b}\\
f_n^{(3)}(X)&=&X^{-(a+n+1)/2}\,{}_2F_1(\fs a\!-\!\fs n,\fs(a\!-\!2b\!+\!1)\!-\!\fs n;\f{3}{2};-X^{-1})\nonumber
\end{eqnarray}
and
\[G_1(n)=\frac{2(2n+1)(n+b)}{(a+n-1)(a+n-2)},\qquad
G_2(n)=\frac{2^{a+n-1}\pi}{\sin \pi(\frac{a+n}{2})}\,\frac{\g(n+\f{3}{2}) \g(n+b+1)}{\g(n+a) \g(\f{3}{2}+\frac{n-a}{2}) \g(b+1+\frac{n-a}{2})},\]
\bee\label{e26c}
G_3(n)=-\frac{2^{a+n}\pi}{\cos \pi(\frac{a+n}{2})}\,\frac{\g(n+\f{3}{2}) \g(n+b+1)}{\g(n+a) \g(1+\frac{n-a}{2}) \g(b+\fs+\frac{n-a}{2})}~.
\ee
The expansion (\ref{e26a}) holds provided $a\neq 0, 1, 2$; that is, not on the upper or lower boundaries of Fig.~1 nor
on the line $d=\fs m+3$ corresponding to $\epsilon=1$ by (\ref{e110}). In these cases, logarithmic terms will be present in (\ref{e26a}).
The coefficient $A_{D,m}$ and the parameters $a$ and $b$ are defined in (\ref{e23})--(\ref{e240}).
\end{theorem}

The special case $a=1$ is considered in Sec. 4.2 (see (\ref{e385a})), and
also in Appendix A where a limiting procedure is employed.

The continuation of the hypergeometric function ${}_3F_2(-X)$ in (\ref{e25}) into $X>1$, represented by the inner series in (\ref{e26a}), possesses a slow monotonic character as $n\to\infty$. From the large-$n$ behaviour of the coefficients ${\cal B}_n$ in (\ref{eBn}), it is seen that the series (\ref{e26a}) converges when $X>1$ provided $\epsilon>0$. However, this convergence is slow on account of the algebraic decay of ${\cal B}_n$.
It is significant that individual series in (\ref{e26a}) (corresponding to separate $j$-values) converge only when $X>X_*$, where $X_*$ will be determined below. It is important to stress, however, that when $1<X\leq X_*$ the divergencies present in these individual series cancel
in the linear combination in (\ref{e26a}) to leave a convergent series.

To determine $X_*$ we examine the large-$n$ behaviour of the two terms on the right-hand side of (\ref{e26a}) corresponding to $j=2, 3$ and define
\bee\label{e382}
h_n(X):=G_2(n) f_n^{(2)}(X)+G_3(n) f_n^{(3)}(X),
\ee
where we suppose that the parameter $a\in (0,2)$ with $a\neq 1$.
Straightforward application of Stirling's approximation for the gamma function shows that
\[G_2(n)\sim \frac{2^b\sqrt{\pi}}{\sin \pi(\frac{a+n}{2})}\, n^{1/2} 4^n,\qquad G_3(n)\sim -\frac{2^b\sqrt{\pi}}{\cos \pi(\frac{a+n}{2})}\, n^{3/2} 4^n\qquad (n\to\infty).\]
We apply Pfaff's transformation \cite[(15.8.1)]{DLMF}
\bee\label{e30}
{}_2F_1(\alpha,\beta;\gamma,z)=(1-z)^{-\alpha}\,{}_2F_1(\alpha,\gamma-\beta;\gamma;z/(z-1))
\ee
to the hypergeometric functions in $f_n^{(j)}(X)$ ($j=2, 3$) to find, with $\zeta=X/(1+X)$,
\[f_n^{(2)}(X)=\frac{\zeta^{(a\!-\!2b\!-\!n)/2}}{X^{(a\!+\!n)/2}}\, {}_2F_1\bl(\frac{a\!-\!2b}{2}\!-\!\frac{1}{2} n,\frac{2\!-\!a}{2}\!+\!\frac{1}{2} n; \frac{1}{2};\frac{1}{1+X}\br),\]
\[f_n^{(3)}(X)=\frac{\zeta^{(a\!-\!2b\!-\!n\!+\!1)/2}}{X^{(a\!+\!n\!+\!1)/2}}\, {}_2F_1\bl(\frac{a\!-\!2b\!+\!1}{2}\!-\!\frac{1}{2} n,\frac{3\!-\!a}{2}\!+\!\frac{1}{2} n; \frac{3}{2};\frac{1}{1+X}\br).\]
The large-$\lambda$ behaviour of Gauss hypergeometric functions of the type ${}_2F_1(\alpha-\lambda,\beta+\lambda;\gamma;z)$ is discussed in \cite[Section 4]{Paris13},
from which it is found with some effort that
\[f_n^{(2)}(X)\sim \frac{\zeta^\mu}{X^{(a+n)/2}} \cos 2\mu\vartheta,\qquad f_n^{(3)}(X)\sim -\frac{1}{n}\,\frac{\zeta^\mu}{X^{(a+n)/2}} \sin 2\mu\vartheta \qquad (n\to \infty),\]
where \[\mu=\fs(a-b-1-n),\qquad \vartheta=\arctan\,X^{-1/2}.\]
Some straightforward algebra then shows that
\[h_n(X)\sim \frac{2^{b-1}\sqrt{\pi}}{\sin \pi a}\,\frac{(-)^n n^{1/2}4^n \zeta^\mu}{X^{(a+n)/2}}\,\cos\bl[2\mu\vartheta-\pi\bl(\frac{a+n}{2}\br)\br]\qquad (n\to\infty).\]

The late behaviour of the terms in the sum $\sum_{n\geq0}{\cal B}_n h_n(X)$ is therefore controlled in modulus by
\[n^{-1/2-\epsilon}\bl(\frac{16}{X\zeta}\br)^{\!n/2},\]
so that the sum converges absolutely if $16/(X\zeta)<1$; that is, if
\bee\label{e380}
X>X_*=4(2+\surd 5)\doteq  16.9443.
\ee
It follows from the above discussion (although we do not examine this here) that the first sum on the right-hand side in (\ref{e26a}), namely $\sum_{n\geq0}{\cal B}_n G_1(n) f_n^{(1)}(X)$, must also converge when $X>X_*$ to yield the finite overall result.

We now establish that the sum on the right-hand side of (\ref{e26a})
corresponding to $j=1$ vanishes for $X>X_*$ when $a\in (0,2)$, $a\neq 1$.
Upon expansion of the ${}_3F_2(-1/X)$ function in $f_n^{(1)}(X)$ as a series in powers of $1/X$, some routine algebra shows that
\bee\label{e57}
\sum_{n\geq0}{\cal B}_n G_1(n) f_n^{(1)}(X)=\frac{2b X^{-1}}{\g(a)}\sum_{k\geq0}\bl(-\frac{4}{X}\br)^k
\sum_{n\geq0}\frac{\g(a\!+\!n\!-\!2\!-\!2k)}{n!}\,(-n+\fs)_k (1-b-n)_k.
\ee
The factor $(-n+\fs)_k (1-b-n)_k$ is a polynomial in $n$ of degree $2k$, which can therefore be written in the form
\[(-n+\fs)_k (1-b-n)_k=\gamma_0+\gamma_1 n+\gamma_2 n(n-1)+ \ldots +\gamma_{2k}n(n-1)\ldots (n+2k-1),\]
where the $\gamma_j$ ($1\leq j\leq 2k$) are computable constants. The inner sum over $n$ in (\ref{e57}) therefore becomes
\[\sum_{j=1}^{2k} \gamma_j \sum_{n\geq j}\frac{\g(a\!+\!n\!-\!2\!-\!2k)}{(n-j)!}=
\sum_{j=1}^{2k} \gamma_j \sum_{n\geq 0}\frac{\g(a\!+\!n\!-\!2\!+\!j\!-\!2k)}{n!}=0\qquad (a<2;\ a\neq 1),\]
since, when $\beta<0$ (cf. \cite[5.2.11.16]{PBM1}),
\[\lim_{x\to 1}\frac{1}{\g(\beta)}\sum_{n\geq0}\frac{\g(n+\beta)}{n!}\,x^n=\lim_{x\to 1}\,{}_1F_0(\beta;;x)=\lim_{x\to 1} (1-x)^{-\beta}=0.\]

Hence, we have the following modified form of the expansion in Theorem \ref{TF}
and its analytic continuation in Theorem \ref{TG}:
\begin{theorem}\label{TH}
$\!\!\!.$\ When $X>X_*=4(2+\surd 5)$ and $a=2-\epsilon\in (0,2)$ $($with $a\neq 1)$ we have the modified expansion
\bee\label{e381}
I_{D,m}(p,q)=A_{D,m}q^{-4+2\epsilon} \sum_{n\geq0} {\cal B}_n h_n(X),
\ee
where ${\cal B}_n$ and $h_n(X)$ are defined in (\ref{b24}) and (\ref{e382}), with $A_{D,m}$ given in (\ref{e23}).
\end{theorem}
Convergence of the sum in (\ref{e381}) is more rapid than that in (\ref{e26a}) due to the presence of the factor $(16/(X\zeta))^{n/2}$, where $16/(X\zeta)<1$ when $X>X_*$,
in the large-$n$ behaviour of $h_n(X)$.

In fact, the result of Theorem \ref{TH} represents the full asymptotic expansion
of $I_{D,m}(p,q)$ as $X\to\infty$ under the condition that $X$ exceeds the ``critical value"
$X_*$. In the next section we give explicit expressions for the
three leading terms of this asymptotic expansion.
As we shall see, in doing this we will also be able to write down the result for
the case $a=1$ prohibited by conditions of the analytic continuation (\ref{ea1})
and of Theorems \ref{TG} and \ref{TH}.

\vspace{0.3cm}

\noindent{\bf  4.2.\ \ Leading terms of the asymptotic expansion for $X\to\infty$}
\vspace{0.2cm}

The leading terms of the asymptotic expansion for $I_{D,m}(p,q)$
as $X\to\infty$ can be derived from (\ref{e381}) as follows.
The hypergeometric functions
appearing in $f_n^{(2, 3)}(X)$
(see (\ref{e26b})) are $1+O(X^{-1})$ as $X\to\infty$.
Thus, with
\bee\label{e382c}
f_0^{(2)}(X)=X^{-a/2} \bl(1-\frac{\beta}{2X}+O(X^{-2})\br),\qquad \beta=(a-1)(a-2b),
\ee
we see from (\ref{e382}) and (\ref{e381}) that
\[I_{D,m}(p,q)= \frac{A_{D,m} q^{-4+2\epsilon}}{X^{a/2}}\bl\{{\cal B}_0G_2(0)+
[{\cal B}_1G_2(1)+{\cal B}_0G_3(0)]X^{-1/2}\hspace{4cm}\]
\bee\label{e382a}
\hspace{4cm}+[{\cal B}_2G_2(2)+{\cal B}_1G_3(1)-\fs\beta {\cal B}_0G_2(0)]X^{-1}+O(X^{-3/2})\br\}
\ee
as $X\to\infty$.
Here, recalling that $a=2-\epsilon$ and $X=4p^2/q^4$, we have
$q^{-4+2\epsilon}/X^{a/2}=(2p)^{-2+\epsilon}$. This means that extracting the main
power of $X$ in the asymptotic expansion and combining it with the
overall factor $q^{-4+2\epsilon}$ leads us to the scaled form
(\ref{e12}) of $I_{D,m}(p,q)$, as it really should be.

Further, at $q=0$ the functional form (\ref{e12}) reduces to
$p^{-2+\epsilon}I_{D,m}(1,0)$ given in (\ref{e13}), with the known
constant $I_{D,m}(1,0)$ shown in (\ref{e14a}).
Using (\ref{e23}) for $A_{D,m}$, (\ref{b24}) for ${\cal B}_0$, and
(\ref{e26c}) for $G_2(0)$, we see indeed after some algebra that
$2^{-2+\epsilon}A_{D,m}{\cal B}_0G_2(0)=I_{D,m}(1,0)$,
which provides a useful check of our calculations.
Hence we obtain the asymptotic expansion in the form
\bee\label{e385}
I_{D,m}(p,q)=p^{-2+\epsilon}I_{D,m}(1,0)\{1+C_1 X^{-1/2}+C_2X^{-1}+O(X^{-3/2})\}
\ee
as $X\to\infty$ when $a\in (0,2)$, where
\[
C_1=\frac{{\cal B}_1G_2(1)+{\cal B}_0G_3(0)}{{\cal B}_0G_2(0)},\qquad
C_2=\frac{{\cal B}_2G_2(2)+{\cal B}_1G_3(1)}{{\cal B}_0G_2(0)}-\frac12\beta,
\]
and $I_{D,m}(1,0)$ is given in (\ref{e14a}).

Using the values
\bee\label{e382b}
{\cal B}_0=1,\qquad {\cal B}_1=\frac{ab}{3(b+1)},\qquad {\cal B}_2=\frac{a(a+1)b}{10(b+2)}
\ee
together with (\ref{e26c}) for the coefficients $G_j(n)$,
we find after some straightforward algebra that
\[
C_1=(a-1)\tan\!\frac{\pi a}2\,\frac{\Gamma(\frac32-\fs a)}{\Gamma(2-\fs a)}
\frac{\Gamma(b+2-\fs a)}{\Gamma(b+\frac32-\fs a)},
\]
\[
C_2=\frac{2b}{3-a}\Big(3\,\frac{b+1-a}{2b+2-a}-a\Big)-\frac12(1-a)(2b-a).
\]

In the particular case $a=1$ ($\epsilon=1$), the coefficient $C_1$ has a finite value
while $C_2$ is regular. These are given by
\bee\label{e385a}
C_1=-\frac{4\Gamma(b+\f32)}{\pi^{3/2}\Gamma(b+1)},\qquad
C_2=\frac{b(b-1)}{2b+1}.
\ee
Although the expansion (\ref{e381}) has not been established when $a=1$, the result (\ref{e385}) with $C_1$ given in (\ref{e385a}) must hold when $a=1$ by appeal to continuity. An alternative derivation of the expansion when $a=1$, which does not assume the result in (\ref{e381}), confirms the expansion (\ref{e385}) in this particular case; see Appendix A for details.

\vspace{0.6cm}

\noindent{\bf  4.3.\ \ The expansion (\ref{e26}) and its modified form}
\vspace{0.2cm}

In order to derive an analytic continuation of the
power series expansion of Theorem \ref{TT}, given by (\ref{e26}),
we employ the technique described in \cite[\S38, pp.~58--59]{Rainville}.
This consists of exploiting the binomial expansion
of $(1-z)^{-\alpha}$ appearing on the right-hand side of Pfaff's
transformation (\ref{e30}) to derive the simple Gauss function
on the left of this relation.

We observed that the transformation (\ref{e30}) can be derived as presented (that is, from left to right) by using the following version of the binomial expansion:
for integer $k$ and $|\zeta|<1$
\[
(1-\zeta)^{-\alpha-k}=\sum_{r\geq0}\frac{(\alpha+k)_r}{r!}\,\zeta^r=
\frac1{(\alpha)_k}\sum_{r\geq0}\frac{(\alpha)_{k+r}}{r!}\,\zeta^r=
\frac{(-)^k}{(\alpha)_k\zeta^k}\sum_{r\geq k}\frac{(-r)_k(\alpha)_r}{r!}\,\zeta^r.
\]
Here we have used the fact that $(\alpha+k)_r=(\alpha)_{k+r}/(\alpha)_k$,
shifted the summation index $r\mapsto r-k$, and taken into account that
$1/(r-k)!=(-)^k(-r)_k/r!$. If we now identify $\zeta$ with $X/(1+X)$  we obtain
the following
\newtheorem{lemma}{Lemma}
\begin{lemma}$\!\!\!.$\ Let $X>0$ and $k$ be a non-negative integer.
Then
\bee\label{e383}
(-X)^k=\frac{(1+X)^{-\alpha}}{(\alpha)_k} \sum_{r\geq k} \frac{(-r)_k (\alpha)_r}{r!}\,\zeta^r,\qquad \zeta=\frac{X}{1+X}\qquad (X>0),
\ee
valid for arbitrary $\alpha>0$.
\end{lemma}
If we use the representation (\ref{e383}) with $X=-z$
in the power expansion of the
Gauss function on the left-hand side of
(\ref{e30}) and change the order of summations,
we easily reproduce the right-hand side of this equation.

Application of Lemma 1 to the series expansion of the $_3F_2$ function
reproduces the corresponding version of N{\o}rlund's formula \cite[(1.21)]{Nor55}
cited in \cite[(16.10.2)]{DLMF},
\bee\label{c0}
{}_3F_2\bl(\begin{array}{c}\alpha, \beta, \gamma\\ \delta, \eta\end{array}\!\!;-x\br)=(1+x)^{-\alpha} \sum_{k\geq 0} \frac{(\alpha)_k}{k!}\,{}_3F_2\bl(\begin{array}{c}-k, \beta, \gamma\\ \delta, \eta\end{array}\!\!;1\br)\,
\bl(\frac{x}{1+x}\br)^k\qquad(x>0).
\ee
Using this formula with the choice $\alpha=a$ in (\ref{e25}), we produce the following result\footnote{
Two other alternative versions of the representation (\ref{c1})--(\ref{c2}) can be obtained from
(\ref{e25}) if we apply there (\ref{c0}) with $\alpha=(a+n)/2$ and $\alpha=(a+n+1)/2$.}
 for our integral $I_{D,m}(X)$.
\begin{theorem}\label{TD}$\!\!\!.$\ For $X>0$ we have the expansion
\bee\label{c1}
I_{D,m}(X)=\frac{A_{D,m}\,q^{-4+2\epsilon}}{1+X}\,\sum_{n,k\geq0} {\cal B}_n F_{nk} \zeta^k,\qquad \zeta=\frac{X}{1+X},
\ee
where the coefficients ${\cal B}_n$ are defined in (\ref{b24}) and $A_{D,m}$ in (\ref{e23}).
The coefficients $F_{nk}$ denote the polynomials
\bee\label{c2}
F_{nk}:={}_3F_2\bl(\begin{array}{c}-k, \frac12 (a+n), \frac12 (a+n+1)\\\f{3}{2}+n, b+1+n\end{array}\!\!;1\br),
\ee
where the parameters $a$ and $b$ are specified in (\ref{e240}).
\end{theorem}
In Appendix D it is shown that inside the convergence domain of Fig.~1 the polynomials $F_{nk}$ satisfy $0<F_{nk}\leq 1$ for non-negative integers $n$ and $k$.

Finally, insertion of the representation (\ref{e383}) into (\ref{e26}) yields
\begin{eqnarray}
I_{D,m}(p,q)&=&\frac{A_{D,m}q^{-4+2\epsilon}}{(1+X)^\alpha} \sum_{k=0}^\infty \frac{{\cal C}_k F_k}{(\alpha)_k} \sum_{r=k}^\infty \frac{(-r)_k (\alpha)_r}{r!}\,\zeta^r\nonumber\\
&=&\frac{A_{D,m}q^{-4+2\epsilon}}{(1+X)^\alpha} \sum_{r=0}^\infty\frac{(\alpha)_r \zeta^r}{r!}\sum_{k=0}^r (-r)_k \frac{{\cal C}_k F_k}{(\alpha)_k}.\label{e384}
\end{eqnarray}
The result (\ref{e384}) has been established assuming $X<1$, but the identity (\ref{e383}) holds for $X>0$ and so
(\ref{e384}) holds by analytic continuation for $X>0$. We now make the choice $\alpha=\fs a$ to accord with the asymptotic estimate in (\ref{e382a}) as $X\to\infty$. Then we have
\begin{theorem}\label{TR}
$\!\!\!.$\ For $X>0$ and $\zeta=X/(1+X)$ we have the expansion
\bee\label{e386}
I_{D,m}(p,q)=\frac{A_{D,m} q^{-4+2\epsilon}}{(1+X)^{a/2}}
\sum_{k=0}^\infty \frac{(\fs a)_k}{k!}\,\Upsilon_k\,\zeta^k,
\qquad \Upsilon_k:=\sum_{r=0}^k (-k)_r\,\frac{{\cal C}_rF_r}{(\fs a)_r},
\ee
where ${\cal C}_r$ and $F_r$ are defined in (\ref{c24}), with $A_{D,m}$, $a$ and $b$ given in (\ref{e23})
and (\ref{e240}).
\end{theorem}
This expansion and that in (\ref{e381}) are tested numerically in Section 7.

\vspace{0.3cm}

\begin{center}
{\bf 5. \  Special cases corresponding to integer $D$}
\end{center}
\setcounter{section}{5}
\setcounter{equation}{0}
\renewcommand{\theequation}{\arabic{section}.\arabic{equation}}
In this section we present special cases of the representations obtained in Section 3 when the dimension $D=1, 2$ and 3.
\vspace{0.3cm}

\noindent{\bf  5.1.\ \ The case $D=3$}
\vspace{0.2cm}

We first consider the simplest situation corresponding to $D=3$.
In this case we have from (\ref{e240}) $b=0$,
so that the coefficients ${\cal B}_n$ in (\ref{b24})
satisfy ${\cal B}_0=1$ and ${\cal B}_n=0$ ($n\geq 1$).
Hence from the zeroth term of (\ref{e25}),
where the $_3F_2(-X)$ with $b=n=0$ contracts to a Gauss hypergeometric function, we obtain
\bee\label{r30}
I_{3,m}(p,q)=A_{3,m} q^{-4+2\epsilon}
\,{}_2F_1(1-\fs\epsilon,\f{3}{2}-\fs\epsilon;\f{3}{2};-X).
\ee
Here we have used $a=2-\epsilon$, which is independent of the choice of $D$ and $m$;
from (\ref{e23})--(\ref{e240}), $A_{3,m}=8\g(2-\epsilon)/(16\pi)^\epsilon$.
Application of Pfaff's transformation in (\ref{e30}) then shows that
\bee\label{e31}
I_{3,m}(p,q)=A_{3,m} q^{-4+2\epsilon}\,(1+X)^{-1+\frac\epsilon2}\,
{}_2F_1\Big(1-\frac{\epsilon}{2},\frac{\epsilon}{2}; \frac32;\frac{X}{1+X}\Big)\qquad (X>0).
\ee
This is the result stated in (\ref{e16}).

We notice that $I_{3,m}(p,q)$ can also be expressed in terms of elementary functions as
\bee\label{g16}
I_{3,m}(p,q)=q^{-4+2\epsilon}\,\frac{8\Gamma(1-\epsilon)}{(16\pi)^\epsilon}\,
\frac1{\sqrt X}\,\Im(1-i\sqrt X)^{-1+\epsilon}
\ee
(another version of (\ref{g16}) can be found in \cite[(A.40)]{RDS11}), or
\bee\label{h16}
I_{3,m}(p,q)=q^{-4+2\epsilon}\,\frac{8\Gamma(1-\epsilon)}{(16\pi)^\epsilon}\,
\frac{(1+X)^{(-1+\epsilon)/2}}{\sqrt X}\,
\sin\Big[(1-\epsilon)\arcsin\sqrt{\frac X{1+X}}\Big],
\ee
cf. \cite[(A7)]{MC99}.
These last two representations follow from (\ref{r30}) and (\ref{e31}) by
\cite[7.3.1.107]{PBM3} and \cite[7.3.1.91]{PBM3}, respectively.
\vspace{0.3cm}

\noindent {\bf 5.2.\ \ The case $D=2$, $m=1$}
\vspace{0.2cm}

When $D=2$, $m=1$ we have the parameters $a=\f{3}{2}$, $b=\fs$ and, from (\ref{e23}),
$A_{2,1}=(2\pi)^{-1}$.
We have been unable to prove the equivalence between (\ref{e26}) and (\ref{e17}), but present an expansion
process that supports our claim that these representations are the same.
Then, from (\ref{e26}), we obtain
\[
I_{2,1}(p,q)=\frac{q^{-3}}{2\pi} \sum_{k\geq0} \frac{(\f{3}{2})_{2k}(-X/4)^k}{(\f{3}{2})_k(\f{3}{2})_k}\,{}_3F_2\bl(\begin{array}{c}2k+\f{3}{2},\fs, \fs\\k+ \f{3}{2}, k+\f{3}{2}\end{array}\!\!;1\br)
\]
\[
=\frac{p^{-3/2}\chi^{3/4}}{4} \sum_{k\geq0} (-\chi)^k {}_3F_2\bl(\begin{array}{c}-k,-k,\fs\nonumber\\ 1,1\end{array}\!\!;1\br)
=\frac{p^{-3/2}\chi^{3/4}}{4} \sum_{k\geq0} (-\chi)^k \sum_{n=0}^k \frac{(-k)_n^2 (\fs)_n}{(n!)^3},
\]
where we have defined $\chi:=X/4=p^2/q^4$.
Here we have employed Thomae's transformation \cite[p.~143]{AAR}
\bee\label{eThomae}
{}_3F_2\bl(\begin{array}{c}\alpha, \beta, \gamma\\ \delta, \eta\end{array}\!\!;1\br)=
\frac{\g(\delta)\g(\eta)\g(s)}{\g(\alpha)\g(s+\beta)\g(s+\gamma)}\,
{}_3F_2\bl(\begin{array}{c}\delta-\alpha, \eta-\alpha, s\\ s+\beta, s+\gamma\end{array}\!\!;1\br),
\ee
where $s=\delta{+}\eta{-}\alpha{-}\beta{-}\gamma$
denotes the parametric excess.
This produces the series expansion, valid when $\chi<\f{1}{4}$,
\bee\label{e36}
I_{2,1}(p,q)=\frac{p^{-3/2}\chi^{3/4}}{4}(1-\f{3}{2}\chi+\f{27}{8}\chi^2-
\f{147}{16}\chi^3+\f{3555}{128}\chi^4-\f{22869}{256}\chi^5+\cdots).
\ee

On the other hand, from (\ref{e17}), where
the variable $w$ expressed in terms of $\chi$ is
\[w=\frac{4\chi^3}{(1+\chi)^2(1+4\chi)}=4\chi^3(1-6\chi+27\chi^2-112\chi^3+ \cdots)\qquad (\chi<\f{1}{4}),\]
we obtain
\[I_{2,1}(p,q)=p^{-3/2}\,\frac{\chi^{3/4}}{4}(1-6\chi+27\chi^2- \cdots)^{1/4} \sum_{n\geq0} \frac{(\f{1}{4})_n^2}{(n!)^2}\,w^n.\]
Expansion of this expression in ascending powers of $\chi$,
when $\chi<\f{1}{4}$, with the help of {\it Mathematica} yields
the same result as in (\ref{e36}).
We have shown that there is agreement between these two expansions up to at least the term involving $\chi^{20}$.
\vspace{0.3cm}

\noindent{\bf 5.3.\ \ The case $D=1$}
\vspace{0.2cm}

When $D=1$ we have from (\ref{e23})--(\ref{e240}) $m=2+2\epsilon$, $b=1$ and
$A_{1,m}=8\,\g(2-\epsilon)/(16\pi)^{1+\epsilon}$.
Then, from (\ref{e26})--(\ref{c24}) and the fact that $(2)_k=k! (k+1)$, we obtain
\bee\label{e32}
I_{1,m}(p,q)=A_{1,m}\,q^{-4+2\epsilon}
\sum_{k\geq0} \frac{(a)_{2k}(-X/4)^k}{(\f{3}{2})_k k! (k+1)}\, {}_3F_2\bl(\begin{array}{c} a+2k, \fs, 1\\\f{3}{2}+k, 2+k\end{array}\!\!;1\br).
\ee
We will now show that the above ${}_3F_2(1)$ series can be reduced to a Gauss hypergeometric series.

From Thomae's transformation (\ref{eThomae}) we obtain on letting $\alpha=\fs$, $\beta=1$,
and $\gamma=a+2k$ with $a=2-\epsilon$, that
\[F(k):=\frac{1}{k+1}\,{}_3F_2\bl(\begin{array}{c}a+2k, \fs, 1\\ \f{3}{2}+k, 2+k\end{array}\!\!;1\br)=2^{-2k-1}\epsilon^{-1}\,{}_3F_2\bl(\begin{array}{c}1+k, \f{3}{2}+k, \epsilon\\ 1+\epsilon, 2+2k\end{array}\!\!;1\br).\]
The parameters in this last hypergeometric series are such that we can apply the quadratic transformation given by \cite[(7.4.1.12)]{PBM3},
\[{}_3F_2\bl(\begin{array}{c}\alpha, \alpha+\fs, \beta\\ \beta+1, 2\alpha\end{array}\!\!;z\br)=Z^{\,2\beta}\,{}_2F_1(\beta, 1+2\beta-2\alpha;\beta+1;1-Z),\qquad Z=\frac{2(1-\sqrt{1-z})}{z},\]
taken at $z=1$. This leads to
\[F(k)=\frac{2^{2\epsilon-1}}{4^k \epsilon}\,{}_2F_1(\epsilon, 2\epsilon-2k-1;1+\epsilon;-1)=
\frac{2^{\epsilon-1}}{4^k \epsilon}\,{}_2F_1(a+2k, \epsilon;1+\epsilon;\fs)\]
upon application of Pfaff's transformation in (\ref{e30}).
It then follows from (\ref{e32}) that
\bee\label{e33}
I_{1,m}(p,q)=\frac{A_{1,m}\,q^{-4+2\epsilon}}{2^{1-\epsilon}\epsilon}\,\sum_{k\geq0} \frac{(a)_{2k}(-X)^k}{(\f{3}{2})_k k!4^{2k}}\,{}_2F_1(a+2k,\epsilon;1+\epsilon;\fs).
\ee
Further implications of this series expansion are considered in the following sub-sections.

\vspace{0.3cm}

\noindent{\bf 5.3.1.\ \ Expression as a Horn function}
\vspace{0.2cm}

In fact, the sum in (\ref{e33}) can be expressed in terms of the Horn function $H_4$ by using
its series representation in terms of Gauss functions \cite[(5.2)]{SP16}, namely
\bee\label{HD}
H_4(\alpha,\beta;\gamma,\delta;x,y)=\sum_{k\geq 0}\frac{(\alpha)_{2k}}{(\gamma)_k\,k!}\;x^k
\,_2F_1\left(\alpha+2k,\beta;\delta;y\right) \qquad (2\sqrt{|x|}+|y|<1),
\ee
while the standard definition of the Horn function $H_4$ as a double series expansion is
\cite[p.~225]{EMOT1}, \cite[p.~57]{SriMan}, \cite[p.~24]{SriKar}
\bee\label{H4D}
H_4(\alpha,\beta;\gamma,\delta;x,y)=\sum_{n,k\geq0}\frac{(\alpha)_{2k+n}(\beta)_n}{(\gamma)_k (\delta)_n}\,\frac{x^k}{k!} \frac{y^n}{n!} \qquad (2\sqrt{|x|}+|y|<1).
\ee
Thus, recalling that $X=4p^2/q^4$ and $a=2-\epsilon$, we obtain the following alternative representation given by
\begin{theorem}$\!\!\!.$\ When $D=1$, we have the representation
\bee\label{e34}
I_{1,m}(p,q)=q^{-4+2\epsilon}\,\frac{2\,\g(2-\epsilon)}{\epsilon\,(8\pi)^{1+\epsilon}}
\,H_4\Big(2-\epsilon,\epsilon;\frac{3}{2},1+\epsilon; \frac{-p^2}{4q^4},\frac{1}{2}\Big)
\qquad (4p^2/q^4<1),
\ee
where $\epsilon$ is defined in (\ref{e110}).
\end{theorem}

The result (\ref{e34}) corrects the similar formula for $I_{1,m}(p,q)$ given
in \cite[Corollary 5.2]{SP16} in terms of the Horn function $H_4$ with arguments $x=-p^2/q^4$ and $y=-1$.
The problem of this latter function is that, according to (\ref{H4D}),
its region of convergence appears to be $p^2/q^4<0$, which is evidently impossible.
The $H_4$ function in \cite[Corollary 5.2]{SP16}, as well as the two $H_4$ functions appearing
at the end of Sec. 5.1 of \cite{SP16}, diverge for any $p,q>0$.

\vspace{0.3cm}

\noindent {\bf 5.3.2.\ \ The complex expansion of $_2F_1$ and the Horn function}
\vspace{0.2cm}

We now demonstrate that the representation (\ref{e34}) in terms of the Horn function $H_4$
is equivalent to the expression in (\ref{e15}) involving the imaginary part of
a Gauss hypergeometric function.
From \cite[(6.8.1.17)]{PBM3} we have
\bee\label{H1}
W:=\Im(1-it)^{-\alpha}{}_2F_1\Big(\alpha,\beta;\gamma;\frac{\sigma}{1-it}\Big)=
\Im\sum_{k\geq 0}\frac{(\alpha)_k}{k!}\,{}_2F_1(\alpha+k, \beta;\gamma;\sigma) (it)^k
\ee
when $|t|<1$ and $|\sigma|<1$. Then, separating the odd terms in the above sum
and taking into account that $(\alpha)_{2k+1}=\alpha(\alpha+1)_{2k}$ and
$(2k+1)!=(\f32)_k k!4^k$, we obtain
\bee\label{H2}
W=\alpha t\sum_{k\geq0}\frac{(\alpha+1)_{2k}}{(\f32)_k\,k!}(-t^2/4)^k
 {}_2F_1(\alpha+1+2k,\beta;\gamma;\sigma).
\ee
The last sum matches the representation (\ref{HD}) of the Horn function
\[
H_4\big(\alpha+1,\beta;\f32,\gamma;-\f14 t^2,\sigma\big)
\]
whose domain of convergence is given by $|t|+|\sigma|<1$.%
\footnote{Until this point, the present calculation has gone in parallel
with that of \cite[p. 546]{SP16}. The formulas \cite[(5.6)]{SP16} and that at the bottom of
\cite[p. 546]{SP16} are appropriate for this region of the variables on matching the notation.}
However, the value $\sigma=-1$, which would be needed to match the Gauss
function in (\ref{e15}) with that of (\ref{H1}), is forbidden by the last condition.
To avoid this difficulty we apply the transformation
(\ref{e30}) to analytically continue the $_2F_1$ function in (\ref{H2}).
Hence we obtain
\[
W=\frac{\alpha t}{(1-\sigma)^{1+\alpha}}
\sum_{k\geq0}\frac{(\alpha+1)_{2k}}{(\f32)_k\,k!}\Big[\frac{-t^2}{4(1-\sigma)^2}\Big]^k
{}_2F_1\Big(\alpha+1+2k,\gamma-\beta;\frac32;\frac{\sigma}{\sigma-1}\Big).
\]
Now the Gauss series converges when $|\sigma/(\sigma-1)|<1$, which implies
$\sigma<1/2$ for real $\sigma$, including the needed value of $\sigma=-1$.
Expressing the last sum in terms of the Horn function $H_4$ we obtain
\bee
\Im(1-it)^{-\alpha}{}_2F_1\Big(\alpha,\beta;\gamma;\frac{\sigma}{1-it}\Big)=
\frac{\alpha\,t}{(1-\sigma)^{1+\alpha}}
H_4\Big(\alpha+1,\gamma-\beta;\frac32,\gamma;
\frac{-t^2}{4(1-\sigma)^2},\frac{\sigma}{\sigma-1}\Big).
\ee
The new Horn series converges when $|t|+|\sigma|<|1-\sigma|$.
This time the value $\sigma=-1$ is allowed, while $|t|<1$ is then required for convergence.
Thus with $\sigma=-1$ we find
\[
\Im(1-it)^{-\alpha}{}_2F_1\Big(\alpha,\beta;\gamma;\frac{-1}{1-it}\Big)=
\frac{\alpha\,t}{2^{1+\alpha}}H_4\Big(\alpha+1,\gamma-\beta;\frac32,\gamma;
\frac{-t^2}{16},\frac12\Big)\qquad(|t|<1).
\]
Substitution of this last result (with $t=\sqrt X$, $\alpha=1-\epsilon$, $\beta=1$,
and $\gamma=1+\epsilon$) into (\ref{e15}) then yields (\ref{e34}),
thereby establishing the equivalence between these two different representations.

\vspace{0.3cm}

\begin{center}
{\bf 6. \  The behaviour of $I_{D,m}(p,q)$ near the upper and lower boundaries}
\end{center}
\setcounter{section}{6}
\setcounter{equation}{0}
\renewcommand{\theequation}{\arabic{section}.\arabic{equation}}
In this section we examine the behaviour of the integral $I_{D,m}(p,q)$ in (\ref{e11}) in the neighbourhood of the lower (see also Appendix B) and upper boundaries of the convergence domain. Inspection of Fig.~1 shows that the lower boundary $d_\ell=\fs m+2$ is encountered for $D$ in the range $0<D<2$, whereas the upper boundary $d_u=\fs m+4$ is encountered in the range $0<D<4$.
In the particular cases $D=1$ and $D=3$, where we use the representations given in (\ref{e15}) and (\ref{e16}), we are able to obtain more precise information on the behaviour of $I_{D,m}(p,q)$ as these boundaries are approached; see Appendix C.


The leading behaviour of $I_{D,m}(p,q)$ when either $p$ or $q$ is zero
near the lower boundary of the convergence domain ($\epsilon\to 0+$) can be obtained from (\ref{e13})--(\ref{e14b}).
Recalling that $m=4-2D+2\epsilon$ and $\epsilon'=2-\epsilon$, we find
\bee\label{e40a}
I_{D,m}(p,0)=\frac{E_D}{\epsilon p^2}+O(1), \qquad I_{D,m}(0,q)=\frac{E_D}{\epsilon q^4}+O(1) \qquad (\epsilon\to 0+),
\ee
where the coefficient $E_D$ is
\bee\label{e40}
E_D=4\frac{(16\pi)^{(D-3)/2}}{\g(\f{3}{2}-\fs D)}~.
\ee

With arbitrary $p$ and $q$, we consider the series representation (\ref{e26}).
Here each ${}_3F_2(1)$ series has a parametric excess $\epsilon$,
so that each of them diverges as $\epsilon\to 0+$ like $1/\epsilon$.
Applying Thomae's transformation in (\ref{eThomae}) to each ${}_3F_2(1)$ series in (\ref{e26}), we obtain
\bee\label{e40b}
{}_3F_2\bl(\begin{array}{c} 2-\epsilon+2k, \fs, b\\ \f{3}{2}+k, b+1+k\end{array}\!\!;1\br)=\frac{\g(\epsilon)\g(\f{3}{2}+k)\g(b+1+k)}{\g(b)\g(\fs+\epsilon)\g(2+2k)}\,
{}_3F_2\bl(\begin{array}{c}\epsilon, 1+k, \f{3}{2}-b+k\\ \fs+\epsilon, 2+2k\end{array}\!\!;1\br).
\ee
This isolates the pole singularity $\sim1/\epsilon$ in the factor $\g(\epsilon)$
and produces a new ${}_3F_2(1)$ series with parametric excess $b=\f{3}{2}-\fs D>0$ when $0<D<2$.
The new ${}_3F_2(1)$ coefficients are $1+O(\epsilon)$ as $\epsilon\to 0+$. Some simple algebra, combining the coefficients ${\cal C}_k$ in (\ref{c24}) with (\ref{e40b}), then yields
\[{\cal C}_k\,{}_3F_2\bl(\begin{array}{c} 2-\epsilon+2k, \fs, b\\ \f{3}{2}+k, b+1+k\end{array}\!\!;1\br)=\frac{4^{-k}b}{2\epsilon}+O(1).\]
Inserting this into (\ref{e26}) and carrying out the summation of the resulting geometric series, we obtain
the behaviour of $I_{D,m}(p,q)$ in the neighbourhood of the lower boundary $d_\ell=\fs m+2$ given by
\bee\label{e41}
I_{D,m}(p,q)=\frac{E_D}{\epsilon(p^2+q^4)}+O(1)\qquad (\epsilon\to 0+),
\ee
where $E_D$ is defined in (\ref{e40}).
This corrects \cite[Eq.~(58)]{SPD05} where the factor $1/(p^2+q^4)$ was omitted. The leading form (\ref{e41}) agrees with the estimates in (\ref{e40a}) and will be examined numerically in Section 7. A heuristic derivation of (\ref{e41}) is given in Appendix B.

The behaviour in the neighbourhood of the upper boundary $d_u=\fs m+4$ is more straightforward. From (\ref{b24})
and the fact that the parameter $a=\epsilon'$, we find as $\epsilon'\to 0+$ that ${\cal B}_0=1$, ${\cal B}_n=O(\epsilon')$ ($n\geq 1$). Hence, from (\ref{e25}), we have in the vicinity of the upper boundary ($\epsilon'\to 0+$)
\[I_{D,m}(p,q)=A_{D,m}\,q^{-2\epsilon'}\,{}_3F_2\bl(\begin{array}{c}1,\fs\epsilon',\fs\epsilon'+\fs\\\!\! \f{3}{2}, \f{5}{2}-\fs D\end{array}\!\!;-X\br)+O(\epsilon')
=A_{D,m}\{1+O(\epsilon')\},\]
since it is readily seen that the ${}_3F_2(-X)$ function is $1+O(\epsilon')$.
From (\ref{e23}), we therefore find (with $m=8-2D-2\epsilon'$)
\bee\label{e42}
I_{D,m}(p,q)=\frac{8\,(16\pi)^{(D-7)/2}}{\epsilon'\,\g(\f{5}{2}-\fs D)}+O(1)
\qquad (\epsilon'\to 0+).
\ee
This can be easily obtained from both (\ref{e14a}) and (\ref{e14b}), as
$\epsilon'\to 0+$, and is in agreement with the pole term in \cite[(39)]{SD01}.

When $D=1, 2$ and 3 this produces the following leading behaviours near the upper boundary
\[I_{1,m}(p,q)\sim\frac{1}{512\pi^3 \epsilon'},\quad I_{2,m}(p,q)\sim\frac{1}{64\pi^3 \epsilon'},\quad
I_{3,m}(p,q)\sim\frac{1}{32\pi^3 \epsilon'}\]
as $\epsilon'\to 0+$.
\begin{figure}[t]
	\begin{center}
{\small($a$)}\includegraphics[width=0.42\textwidth]{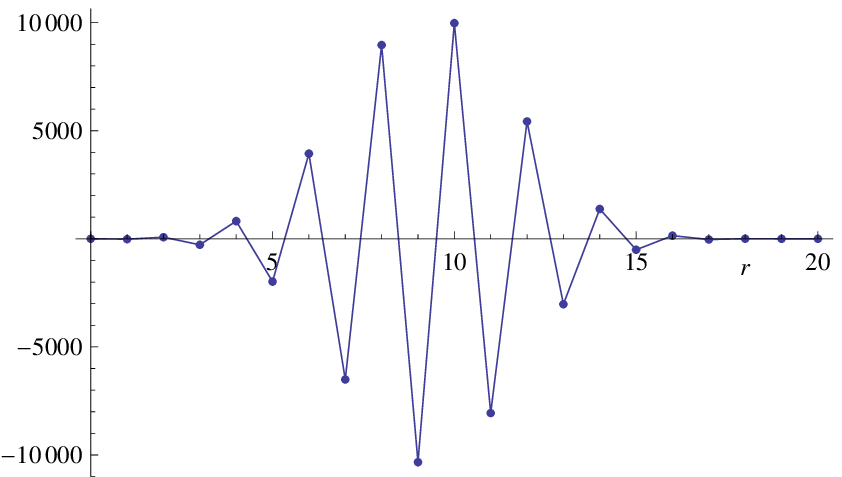}\qquad
{\small($b$)}	\includegraphics[width=0.42\textwidth]{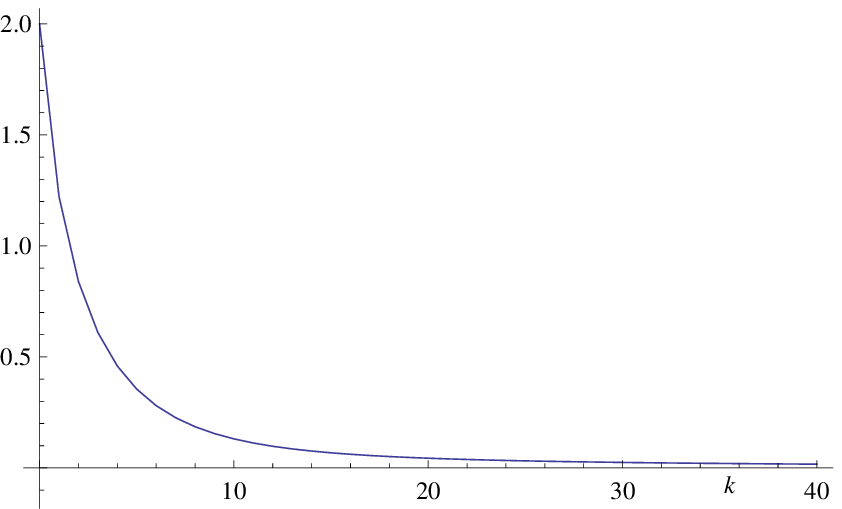}
\caption{\small{(a) The behaviour of the terms in the sum $\Upsilon_k$ when $k=20$ and (b) the behaviour of $\Upsilon_k$ as a function of $k$ when $D=1$, $\epsilon=0.50$. The curves are shown continuous for clarity.}}
	\end{center}
\end{figure}
\vspace{0.4cm}

\vspace{0.6cm}

\begin{center}
{\bf 7. \  Numerical evaluation}
\end{center}
\setcounter{section}{7}
\setcounter{equation}{0}
\renewcommand{\theequation}{\arabic{section}.\arabic{equation}}
In our numerical calculations we shall find it expedient to evaluate the function
\bee\label{e51}
{\hat I}_{D,m}(X):=q^{4-2\epsilon} I_{D,m}(p,q),
\ee
for which the dependence on $p$ and $q$ is solely in terms of the variable $X=4p^2/q^4$. The expansions we employ in the numerical evaluation of ${\hat I}_{D,m}(X)$ are the modified versions of the expansions in Theorems \ref{TH} and \ref{TR} discussed in Section 4.

In Table 1 we show the computed values of ${\hat I}_{D,m}(X)$ when $D=1$ employing the expansions in (\ref{e381}) and (\ref{e386}) compared with the exact evaluation in (\ref{e15}) given by
\begin{table}[t]
\caption{\footnotesize{Values of ${\hat I}_{D,m}(X)$ when $D=1$ for varying values of $X$ and $\epsilon$ obtained from the exact solution in (\ref{e52}) and the series expansions in (\ref{e381}) and (\ref{e386}).}}
\begin{center}
\begin{tabular}{|r||ll|ll|ll|}
\hline
&&&&&&\\[-0.3cm]
\mcol{1}{|c||}{} & \mcol{2}{c|}{$\epsilon=1.8$} & \mcol{2}{c|}{$\epsilon=1.5$} & \mcol{2}{c|}{$\epsilon=1.2$}\\
\mcol{1}{|c||}{$X$} & \mcol{1}{c}{Exact} & \mcol{1}{c|}{Series} & \mcol{1}{c}{Exact} & \mcol{1}{c|}{Series} & \mcol{1}{c}{Exact} & \mcol{1}{c|}{Series}\\
[.1cm]\hline
&&&&&&\\[-0.3cm]
0.5 & 0.00065703 & 0.00065703 & 0.00087461 & 0.00087461 & 0.00201263 & 0.00201263\\
1.0 & 0.00065101 & 0.00065101 & 0.00085127 & 0.00085127 & 0.00191690 & 0.00191690\\
5.0 & 0.00062111 & 0.00062111 & 0.00074463 & 0.00074463 & 0.00151469 & 0.00151469\\
10.0& 0.00060034 & 0.00060034 & 0.00067805 & 0.00067805 & 0.00128944 & 0.00128944\\
15.0& 0.00058629 & 0.00058629 & 0.00063593 & 0.00063592 & 0.00115635 & 0.00115632\\
20.0& 0.00057564 & 0.00057561 & 0.00060540 & 0.00060534 & 0.00106425 & 0.00106411\\
30.0& 0.00055984 & 0.00055984 & 0.00056224 & 0.00056224 & 0.00094018 & 0.00094018\\
50.0& 0.00053900 & 0.00053900 & 0.00050884 & 0.00059884 & 0.00079643 & 0.00079643\\
100.0&0.00050980 & 0.00050980 & 0.00044024 & 0.00044024 & 0.00062729 & 0.00062729\\
[.2cm]\hline
&&&&&&\\[-0.3cm]
\mcol{1}{|c||}{} & \mcol{2}{c|}{$\epsilon=1.0$} & \mcol{2}{c|}{$\epsilon=0.5$} & \mcol{2}{c|}{$\epsilon=0.1$}\\
\mcol{1}{|c||}{$X$} & \mcol{1}{c}{Exact} & \mcol{1}{c|}{Series} & \mcol{1}{c}{Exact} & \mcol{1}{c|}{Series} & \mcol{1}{c}{Exact} & \mcol{1}{c|}{Series}\\
[.1cm]\hline
&&&&&&\\[-0.3cm]
0.5 & 0.00405340 & 0.00405340 & 0.03500095 & 0.03500095 & 0.55518002 & 0.55518002\\
1.0 & 0.00379980 & 0.00379980 & 0.03155789 & 0.03155789 & 0.49522817 & 0.49522817\\
5.0 & 0.00279009 & 0.00279009 & 0.01924598 & 0.01924598 & 0.27407565 & 0.27407565\\
10.0& 0.00226386 & 0.00226385 & 0.01377792 & 0.01377790 & 0.17912771 & 0.17912760\\
15.0& 0.00196682 & 0.00196675 & 0.01101157 & 0.01101125 & 0.13397689 & 0.13397566\\
20.0& 0.00176760 & 0.00176735 & 0.00929419 & 0.00929304 & 0.10739515 & 0.10739091\\
30.0& 0.00150780 & 0.00150780 & 0.00722907 & 0.00722907 & 0.07730222 & 0.07730222\\
50.0& 0.00121983 & 0.00121983 & 0.00518223 & 0.00518223 & 0.04999059 & 0.04999059\\
100.0&0.00090051 & 0.00090051 & 0.00323033 & 0.00323033 & 0.02696565 & 0.02696565\\
[.2cm]\hline
\end{tabular}
\end{center}
\end{table}
\bee\label{e52}
{\hat I}_{1,m}(X)=\frac{\g(1-\epsilon)}{\pi\epsilon(16\pi)^\epsilon}\,
\frac1{\sqrt X}\,\Im \bl\{(1-i\sqrt X)^{-1+\epsilon}
{}_2F_1\bl(1,1-\epsilon;1+\epsilon;\frac{-1}{1-i\sqrt X}\br)\br\}
\ee
for various values of $X$ and $\epsilon$. For $X\leq 20$, we employed the expansion (\ref{e386}) truncated at $k=20$ for small values of $X$ rising to $k=80$ for larger values of $X$.
The terms in $\Upsilon_k$ are found to oscillate (because $(-k)_r=(-)^r k!/(k-r)!$) increasing to a maximum value (in modulus) near $k\simeq r/2$
followed by a steady decrease as $r\to k$. In the sum $\Upsilon_k$ these oscillatory terms largely cancel to yield a slowly decaying function of $k$; see Fig.~2 for a typical example. However, it should be remarked that as $X$ increases ($\zeta\to 1$) there will be a loss of accuracy at fixed precision, since the evaluation of (\ref{e386}) will require larger values of $k$, which in turn results in a more extreme cancellation of terms in $\Upsilon_k$. For this reason, we employed the expansion (\ref{e381}) for $X>20$ with the summation index truncated at $n=20$ for larger $\epsilon$ values rising to $n=80$ as $\epsilon\to 0$.  The tabulated values confirm the accuracy of the expansions (\ref{e381}) and (\ref{e386}). Table 2 presents values of ${\hat I}_{D,m}(X)$ for non-integer $D$ and different values of $X$ and $\epsilon$.
\begin{table}[h]
\caption{\footnotesize{Values of ${\hat I}_{D,m}(X)$ for non-integer $D$ and varying values of $X$ and $\epsilon$ obtained from the series expansions in (\ref{e381}) and (\ref{e386}).}}
\begin{center}
\begin{tabular}{|r||lll|lll|}
\hline
&&&&&&\\[-0.3cm]
\mcol{1}{|c||}{} & \mcol{3}{c|}{$D=0.5$} & \mcol{3}{c|}{$D=1.5$}\\
\mcol{1}{|c||}{$\epsilon$} & \mcol{1}{c}{${\hat I}_{D,m}(1)$} & \mcol{1}{c}{${\hat I}_{D,m}(5)$} & \mcol{1}{c|}{${\hat I}_{D,m}(25)$} & \mcol{1}{c}{${\hat I}_{D,m}(1)$} & \mcol{1}{c}{${\hat I}_{D,m}(5)$} & \mcol{1}{c|}{${\hat I}_{D,m}(25)$}\\
[.1cm]\hline
&&&&&&\\[-0.3cm]
0.1 & 0.19773960 & 0.11047963 & 0.03641803 & 1.13386690 & 0.61870843 & 0.20111954\\
0.5 & 0.01160331 & 0.00720346 & 0.00307534 & 0.08045848 & 0.04792236 & 0.01984903\\
1.0 & 0.00132736 & 0.00098731 & 0.00058022 & 0.01032270 & 0.00745876 & 0.00427713\\
1.2 & 0.00066025 & 0.00052705 & 0.00034947 & 0.00529831 & 0.00413478 & 0.00268688\\
1.5 & 0.00028822 & 0.00025364 & 0.00019931 & 0.00240296 & 0.00208671 & 0.00161876\\
1.8 & 0.00021743 & 0.00020790 & 0.00019027 & 0.00186869 & 0.00177815 & 0.00161892\\
[.2cm]\hline
\end{tabular}
\end{center}
\end{table}

In Fig.~3 we show the behaviour of ${\hat I}_{D,m}(X)$ as a function of $\epsilon$ when $D=\fs$ and $D=\f{3}{2}$
and $X=2$. The dashed curves are the leading approximations as $\epsilon\to 0$ and $\epsilon'\to 0$ obtained from (\ref{e41}) and (\ref{e42}) given by
\[{\hat I}_{D,m}(X)\sim \frac{E_D}{\epsilon (1+X/4)}\quad (\epsilon\to 0),\qquad {\hat I}_{D,m}(X)\sim
\frac{2^{D-4}(4\pi)^{(D-7)/2}}{\epsilon' \g(\f{5}{2}-\fs D)}\quad (\epsilon'\to0).\]
The graphs are split at $\epsilon=1$ due to the different variation in the neighbourhood of $\epsilon=0$ and $\epsilon=2$.
\begin{figure}[h]
	\begin{center}
{\small($a$)}\includegraphics[width=0.42\textwidth]{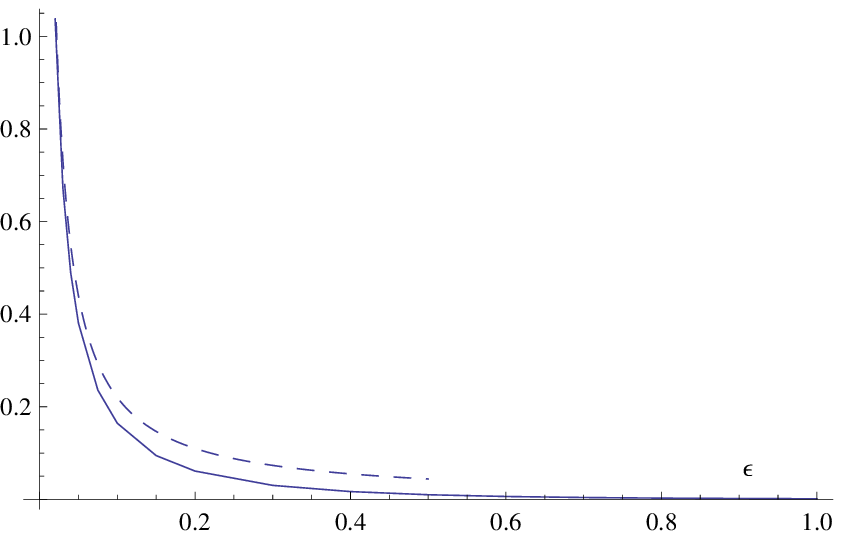}\qquad
	\includegraphics[width=0.42\textwidth]{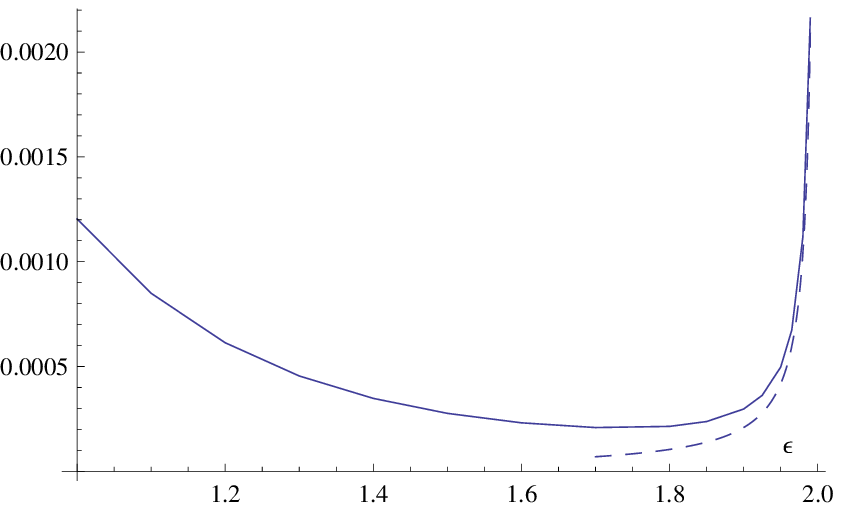}\\
	\vspace{0.3cm}
	
	{\small($b$)}\includegraphics[width=0.42\textwidth]{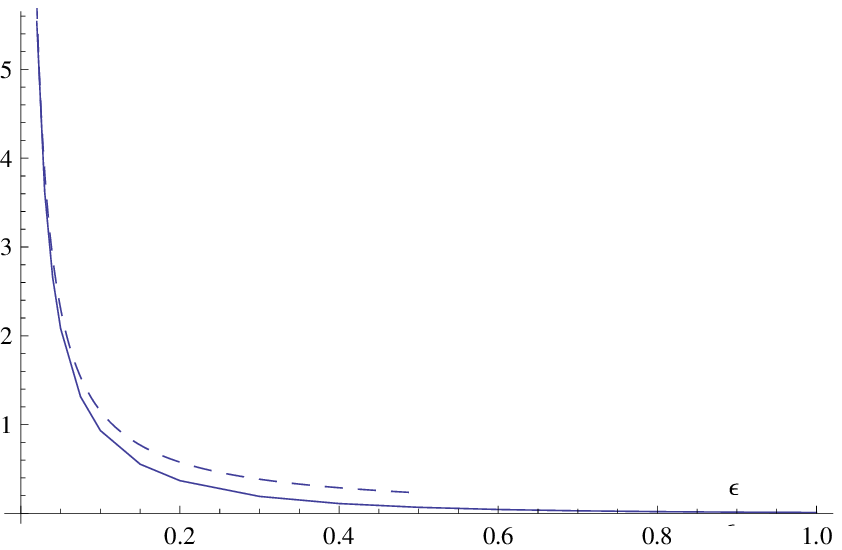}\qquad
	\includegraphics[width=0.42\textwidth]{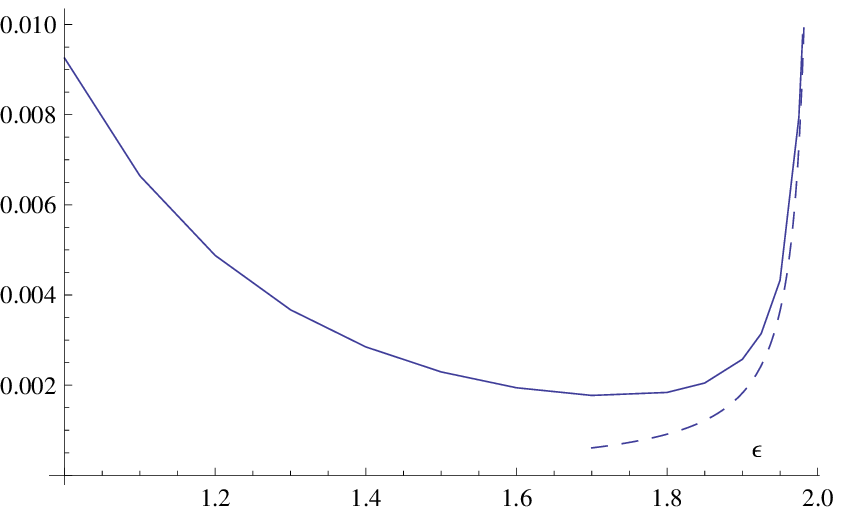}		
\caption{\small{The behaviour of ${\hat I}_{D,m}(X)$ as a function of $\epsilon$ when $X=2$ and ($a$) $D=1/2$ and ($b$) $D=3/2$.}}
	\end{center}
\end{figure}

Finally, in Table 3 we illustrate the accuracy of the asymptotic expansion (\ref{e385}) by presenting values of the absolute relative error in the calculation of ${\hat I}_{D,m}(X)$ compared with the computed values using (\ref{e386}).
\begin{table}[h]
\caption{\footnotesize{Values of the absolute relative error in ${\hat I}_{D,m}(X)$ obtained from the asymptotic expansion in (\ref{e385}).}}
\begin{center}
\begin{tabular}{|r||c|c|c|}
\hline
&&&\\[-0.3cm]
\mcol{1}{|c||}{$X$} & \mcol{1}{c|}{$D=0.5$,\ $\epsilon=1.80$}
& \mcol{1}{c|}{$D=1.5$,\ $\epsilon=1.25$} & \mcol{1}{c|}{$D=2$,\ $\epsilon=0.50$}\\
[.1cm]\hline
&&&\\[-0.3cm]
20 & $1.288\times10^{-2}$ & $5.217\times10^{-3}$ & $2.344\times10^{-2}$\\
40 & $4.354\times10^{-3}$ & $1.844\times10^{-3}$ & $7.521\times10^{-3}$\\
60 & $2.315\times10^{-3}$ & $1.001\times10^{-3}$ & $3.901\times10^{-3}$\\
80 & $1.481\times10^{-3}$ & $6.484\times10^{-4}$ & $2.458\times10^{-3}$\\
100& $1.048\times10^{-3}$ & $4.630\times10^{-4}$ & $1.722\times10^{-3}$\\
150& $5.606\times10^{-4}$ & $2.510\times10^{-4}$ & $9.053\times10^{-4}$\\
[.2cm]\hline
\end{tabular}
\end{center}
\end{table}
Fig.~4 shows the variation of ${\hat I}_{D,m}(X)$ (on a logarithmic scale) as a function of $X$ for varying values of $D$ and a fixed value of $\epsilon$.
\begin{figure}[h]
	\begin{center}
\includegraphics[width=0.50\textwidth]{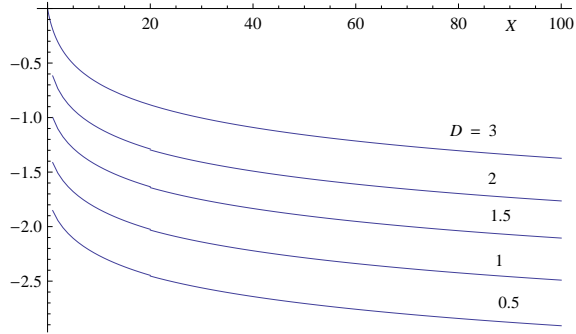}
\caption{\small{The behaviour of $\log_{10} {\hat I}_{D,m}(X)$ as a function of $X$ for varying $D$ when $\epsilon=0.50$.}}
	\end{center}
\end{figure}

\vspace{0.6cm}

\begin{center}
{\bf 8. \  Concluding remarks}
\end{center}
\setcounter{section}{8}
\setcounter{equation}{0}
\renewcommand{\theequation}{\arabic{section}.\arabic{equation}}
We have calculated the one-loop, two-point, massless Feynman integral (\ref{e11})
ubiquitous in the theory of (multi)critical behaviour of strongly
anisotropic systems at Lifshitz points.
The same kind of integral is relevant in the $z=2$ Lorentz-violating
quantum field theories whose origins stem from the above statistical physics problem
(the above value of the parameter $z$ is related to a more general definition
of powers of the modulus of the momentum  $\bm q$ in (\ref{e11}) as $q^{2z}$).

The integral (\ref{e11}) is defined in a $d$ dimensional Euclidean space
composed of two
complementary subspaces of co-dimensions $D$ and $m$ and thus is much more
complicated compared to its usual (massive) counterpart discussed at the beginning of
Section 3.
While the calculation of the latter has attracted attention for decades,
the Feynman integral (\ref{e11}) with general $D$ and $m$ ($D+m=d$, see Fig.1)
is calculated here for the first time.

We have obtained our results first in a form of series expansions
in powers of the variable $X<1$ and in the form of functional series
involving generalised hypergeometric functions $_3F_2$ of the
argument $-X$ with the same constraint $X<1$.
Subsequently, in several possible ways, we have solved a non-trivial mathematical
problem of the analytic continuation of both these series into the region
of arbitrary $X\ge1$. We have derived and discussed the asymptotic expansion
of the integral (\ref{e11}) for large values of $X$.

We have considered all known special cases and possible limiting regimes where
our general results reduce to previously known results.
Yet more confidence in our findings is provided by the numerical section
where numerous tests are carried out in comparison with the analytically solved
case of $D=1$.
A number of graphical presentations provides the reader with certain visualisations.
Several inconsistencies in previous evaluations of $I_{D,m}(p,q)$ have been pointed out.

In a subsequent publication, we plan to use these results to compute the
values of the first-order coefficients of the large-$n$ expansion
of the Lifshitz point's correlation critical exponents $\eta_{L2}$ and $\eta_{L4}$
along several lines, e.g. $D=1$, $D=3$, $m=1$ inside the convergence region
of the Fig. 1 and to produce the corresponding plots.

\vspace{0.6cm}

\begin{center}
{\bf Appendix A: \ The asymptotic expansion for $I_{D,m}(X)$ as $X\to\infty$ when $a=1$}
\end{center}
\setcounter{section}{3}
\setcounter{equation}{0}
\renewcommand{\theequation}{\Alph{section}.\arabic{equation}}
As an alternative to the calculation of Sec. 4.2,
in this appendix we consider Theorem \ref{TG} (see (\ref{e26a})--(\ref{e26c})) and
determine the expansion of $I_{D,m}(p,q)$ as $X\to\infty$ in the particular case $a=1$ ($\epsilon=\epsilon'=1$) when the condition in Theorem \ref{TG} is not applicable.
As in Section 4.2, we shall retain terms up and including $O(X^{-3/2})$. This means that we can ignore the hypergeometric functions
appearing in (\ref{e26b}) since they are $1+O(X^{-1})$ in this limit, with the exception of that in $f_0^{(2)}(X)$ given in (\ref{e382c}). Thus, from (\ref{e26a}), we find
when $X\to\infty$
\[
I_{D,m}(p,q)=A_{D,m}q^{-4+2\epsilon} \{S-\fs\beta{\cal B}_0 G_2(0)X^{-1-a/2}+O(X^{-2},X^{-3/2-a/2})\},\]
where
\[S:=\sum_{n\geq0}{\cal B}_n\bl(\frac{G_1(n)}{X}+\frac{G_2(n)}{X^{(a+n)/2}}+\frac{G_3(n)}{X^{(a+n+1)/2}}\br)\]
and $\beta=(a-1)(a-2b)$.
We now set $a=1+\delta$ and consider the limit $\delta\to 0$. From (\ref{e26c}), the coefficients $G_1(n)=O(\delta^{-1})$ for $n=0, 1$ and $G_1(n)=O(1)$ for $n\geq 2$; also we have $G_2(1)$, $G_3(0)=O(\delta^{-1})$, with $G_2(0)$, $G_2(2)$ and $G_3(1)$ being finite in this limit. Then we obtain
\[
S=\frac{{\cal B}_0 G_2(0)}{X^{(1+\delta)/2}}+\frac{{\cal B}_0}{X}\bl\{G_1(0)+\frac{G_3(0)}{X^{\delta/2}}\br\}+
\frac{{\cal B}_1}{X}\bl\{G_1(1)+\frac{G_2(1)}{X^{\delta/2}}\br\}+
\frac{1}{X}\sum_{n\geq 2}{\cal B}_nG_1(n)\]
\[+\frac{1}{X^{(3+\delta)/2}}\{{\cal B}_2G_2(2)+{\cal B}_1 G_3(1)\}+O(X^{-2-\delta/2}).
\]

From the definition of $G_j(n)$ in (\ref{e26c}), we find the following expansions when $a\to 1$
\[G_1(0)=-\frac{2b}{\delta}(1+\delta+\cdots),\qquad G_1(1)=\frac{6(b+1)}{\delta}(1-\delta+\cdots),\]
\[G_3(0)=\frac{2b}{\delta}(1+{\hat A}\delta+\cdots), \qquad G_2(1)=-\frac{6(b+1)}{\delta}(1+{\hat B}\delta+\cdots),\]
with
\[{\hat A}=-\psi(1)+\fs\psi(\fs)+\fs\psi(b)+\log\,2,\qquad {\hat B}=-\psi(2)+\fs\psi(\f{3}{2})+\fs\psi(b+1)+\log\,2.\]
It then follows that the terms appearing in the $X^{-1}$ contribution are
\[G_1(0)+G_3(0)X^{-\delta/2}=b\{-2+2{\hat A}-\log X\}+O(\delta),\]
\[G_1(1)+G_2(1)X^{-\delta/2}=3(b+1)\{-2-2{\hat B}+\log X\}+O(\delta),\]
and, with the values of ${\cal B}_n$ given in (\ref{b24}) when $a=1$,
\[\sum_{n\geq 2}{\cal B}_nG_1(n)=2b\sum_{n\geq 2}\frac{1}{n(n-1)}=2b.\]
Collecting all terms up to $O(X^{-1})$, we then find when $\delta=0$
\[
S={\cal B}_0\frac{G_2(0)}{X^{1/2}}-\frac{2b}{X}(1-{\hat A}+{\hat B})+\frac{1}{X^{3/2}}\{{\cal B}_2G_2(2)+{\cal B}_1 G_3(1)\}+O(X^{-2}).
\]
Use of $\psi(x+1)-\psi(x)=1/x$ satisfied by the psi-function shows that
${\hat B}-{\hat A}=1/(2b)$.
We observe also that the $\log X$ terms have cancelled in the $O(X^{-1})$ term.

Insertion of the (regular) values of $G_2(0)$, $G_2(2)$ and $G_3(1)$ at $a=1$ obtained from (\ref{e26c}), and of ${\cal B}_0$, ${\cal B}_1$ and ${\cal B}_2$ from (\ref{e382b}), then yields the expansion when $a=1$ ($\epsilon=1$, $\beta=0$)
\bee\label{cc1}
I_{D,m}(p,q)=\frac{(16\pi)^{\f D2-1}\g(b+1)}{32p\,\g(b+\fs)} \bl\{1-\frac{4\g(b+\f32)}{\pi^{3/2}\g(b+1)}X^{-1/2}
+\frac{b(b-1)}{2b+1}X^{-1}+O(X^{-3/2})\br\}
\ee
as $X\to\infty$, where we recall that $b=(3-D)/2$. This agrees with the result obtained by substituting $a=1$ in (\ref{e385}) with the coefficients $C_1$ and $C_2$ given by (\ref{e385a}),
thereby confirming the validity of (\ref{e385}) in the particular case $a=1$
when Theorem \ref{TG} is not applicable.
\vspace{0.6cm}

\begin{center}
{\bf Appendix B: \ Heuristic derivation of (\ref{e41})}
\end{center}
\setcounter{section}{2}
\setcounter{equation}{0}
\renewcommand{\theequation}{\Alph{section}.\arabic{equation}}

In its integration range, the integral (\ref{e11}) encounters convergence problems
when its denominators become infinitesimally small or infinitely large.
In dealing with the current momentum representation, the related singularities
are termed
as infrared and ultraviolet, respectively. In this appendix we consider the
first kind of
problem.

When the integration momenta $\bm p'$ and $\bm q'$ in (\ref{e11}) are close to zero
we can neglect them in the sums $\bm p'+\bm p$ and $\bm q'+\bm q$.
Then the leading contribution to the integral $I_{D,m}(p,q)$ coming from the region
of vanishing $\bm p'$ and $\bm q'$ is approximated by
\[
I_{D,m}^{(0)}(p,q)\simeq\int\frac{d^mq'}{(2\pi)^m}\!\int\frac{d^Dp'}
{(2\pi)^D}\,\frac{1}{p'^2+q'^4}\,\frac{1}{p^2+q^4},
\]
where it is understood that integration variables
are cut off at some finite $\Lambda$.

Similarly, when $\bm p'$ and $\bm q'$ are close to $-\bm p$ and $-\bm q$
we have for the main contribution to (\ref{e11}) in this integration region
\[
I_{D,m}^{(1)}(p,q)\simeq\int\frac{d^mq'}{(2\pi)^m}\!\int\frac{d^Dp'}
{(2\pi)^D}\,\frac{1}{p^2+q^4}\,\frac{1}{(\bm p'+\bm p)^2+(\bm q'+\bm q)^4},
\quad \bm p'+\bm p,\;\bm q'+\bm q\to\bm0.
\]
Application of simple shifts of integration momenta here reduces $I_{D,m}^{(1)}(p,q)$
to $I_{D,m}^{(0)}(p,q)$.

Thus, summing up the above singular contributions leads us to the
approximation
\[
I_{D,m}(p,q)\simeq 2I_{D,m}^{(0)}(p,q)=
\frac2{p^2+q^4}\int\frac{d^my}{(2\pi)^m}\!\int\frac{d^Dx}{(2\pi)^D}\,
\frac{1}{x^2+y^4}\,.
\]
The integral over $x$ converges in the infinite range and is given by
\[\int\frac{d^Dx}{(2\pi)^D}\,\frac{1}{x^2+y^4}={\hat\gamma}_D \,y^{2D-4},\qquad
{\hat\gamma}_D:=\frac{\g(1-\fs D)}{(4\pi)^{D/2}}.\]
Then
\[I_{D,m}(p,q)\simeq\frac{2{\hat\gamma}_D}{p^2+q^4}\int\frac{d^my}{(2\pi)^m}
y^{2D-4}:=
\frac{2{\hat\gamma}_DK_m}{p^2+q^4}\int_0^\Lambda y^{m-1} y^{2D-4}dy\]
\[=\frac{2{\hat\gamma}_DK_m}{p^2+q^4}\frac{\Lambda^{2\epsilon}}{2\epsilon},\]
where $K_m$ is the geometric factor introduced in (\ref{e20}) and, from
(\ref{e110}), $\epsilon=D-2+\fs m>0$. Thus we obtain an $\epsilon$ pole at the
lower boundary and
\[I_{D,m}(p,q)\simeq\frac{1}{\epsilon(p^2+q^4)}\,{\hat\gamma}_DK_m|_{\epsilon=0}+O(1),\]
where
\[{\hat\gamma}_DK_m|_{\epsilon=0}=\frac{\g(1-\fs D)}{(4\pi)^{(d)/2} \g(\fs
m)}\br|_{\epsilon=0}=4\frac{(16\pi)^{(D-3)/2}}{\g(\f{3}{2}-\fs D)}~.\]
This corresponds to the form stated in (\ref{e41}).
\vspace{0.6cm}

\begin{center}
{\bf Appendix C: \ More precise behaviour near the boundaries when $D=1$ and $D=3$}
\end{center}
\setcounter{section}{3}
\setcounter{equation}{0}
\renewcommand{\theequation}{\Alph{section}.\arabic{equation}}
We now examine in more detail the expansion of $I_{D,m}(p,q)$ for $D=1$ near the
lower and upper boundaries (that is, $\epsilon\to0+$ and $\epsilon'\to 0+$)
and for $D=3$ near the upper boundary  ($\epsilon'\to 0+$).
\vspace{0.2cm}

\noindent{\bf C1.\ \ The case $D=1$}
\vspace{0.2cm}

Consider first the case $D=1$, where $m=2+2\epsilon$ and let $\epsilon\to 0+$.
From (\ref{e15}) we obtain
\bee\label{e43b}
I_{1,m}(p,q)=\frac{q^{-4}}{\pi\epsilon \xi}\,\frac{q^{2\epsilon}
\g(1-\epsilon)}{(16\pi)^\epsilon}\,\Im\{z^{1-\epsilon}
{}_2F_1(1,1-\epsilon;1+\epsilon;-z)\},\qquad z:=(1-i\xi)^{-1},
\ee
where for convenience we have set $\xi=X^{1/2}=2p/q^2$. The hypergeometric
function in (\ref{e43b}) has the small-$\epsilon$ expansion given by
\[{}_2F_1(1,1-\epsilon;1+\epsilon;-z)=\frac{\g(1+\epsilon)}{\g(1-\epsilon)}
\sum_{n\geq0}\frac{\g(1-\epsilon+n)}{\g(1+\epsilon+n)}\,(-z)^n=\frac{\g(1+\epsilon)}{\g(1-\epsilon)}\sum_{n\geq0}\{1-2\epsilon
\psi(n+1)+O(\epsilon^2)\}(-z)^n\]
\[=\frac{\g(1+\epsilon)}{\g(1-\epsilon)}\,\frac{1}{1+z}
\bl\{1+2\epsilon\{\gamma+\log\,(1+z)\}+O(\epsilon^2)\br\},\]
where $\psi(x)$ is the logarithmic derivative of the gamma function,
$\gamma=0.57721\ldots$ is the Euler-Mascheroni constant and we note that $|z|<1$
for $\xi>0$.
Then some routine algebra yields
\begin{eqnarray*}
I_{1,m}(p,q)&=&\frac{q^{-4}}{\pi\epsilon \xi(4+X)}\,\frac{q^{2\epsilon}
\g(1+\epsilon)}{(16\pi)^\epsilon}\,\Im
\bl\{(2+i\xi)(1-i\xi)^\epsilon
\bl(1+2\epsilon\bl\{\gamma+\log\,\frac{2-i\xi}{1-i\xi}\br\}+O(\epsilon^2)\br)\br\}\\
&=&\frac{q^{-4}}{\pi\epsilon (4+X)}\,\frac{q^{2\epsilon}
\g(1+\epsilon)}{(16\pi)^\epsilon}\bl\{1+2\epsilon\gamma+\epsilon\log\,
\frac{4+X}{\sqrt{1+X}}+\frac{2\epsilon}{\xi}(\phi-2\omega)+O(\epsilon^2)\br\}
\end{eqnarray*}
where we have introduced the phase angles
\bee\label{e400}
\omega:=\arctan\,(\xi/2),\qquad \phi:=\arctan\,\xi.
\ee
Upon observing that $2\omega-\phi=\arctan\,\xi^3/(4+3\xi^2)$ and
$q^{2\epsilon}\g(1+\epsilon)/(16\pi)^\epsilon=1-\epsilon\{\gamma+\log\,(16\pi/q^2)\}+O(\epsilon^2)$,
we finally find that
\bee\label{e43}
I_{1,m}(p,q)=\frac{1}{4\pi(p^2+q^4)}\bl\{\frac{1}{\epsilon}+\gamma+\log\,\frac{p^2+q^4}{4\pi\sqrt{4p^2+q^4}}-\frac{q^2}{p}\arctan\,\frac{2p^3/q^2}{q^4+3p^2}+O(\epsilon)\br\}
\ee
as $\epsilon\to 0+$. The expression (\ref{e43}) provides a more precise
description in the neighbourhood of the lower boundary when $D=1$.

Now consider the case $D=1$ near the upper boundary, where $m=6-2\epsilon'$, and
let $\epsilon'\to 0+$.
From (\ref{e43b}), with $\epsilon$ replaced by $2-\epsilon'$, we have
\[I_{1,m}(p,q)=\frac{-q^{-2\epsilon'}\g(1+\epsilon')}{\pi\epsilon'(1-\epsilon')(2-\epsilon')
(16\pi)^{2-\epsilon'} \xi}\,\Im\{z^{\epsilon'-1}
{}_2F_1(1,-1+\epsilon';3-\epsilon';-z)\}.\]
Application of \cite[(15.5.13)]{DLMF} to the hypergeometric function shows that
\begin{eqnarray*}
{\cal
F}\equiv{}_2F_1(1,-1+\epsilon';3-\epsilon';-z)&=&\frac{2-\epsilon'}{3-2\epsilon'}+\frac{1-\epsilon'}{3-2\epsilon'}\,(1+z)\,{}_2F_1(\epsilon',1;3-\epsilon';-z)\\
&=&1+\frac{z}{3}-\frac{\epsilon' z}{9}+\frac{1}{3}(1+z) \sum_{n=1}^\infty
\frac{(\epsilon')_n}{(3-\epsilon')_n} (-z)^n +O(\epsilon'\,{}^2)\\
&=&1+\frac{z}{3}-\frac{\epsilon' z}{9}-\frac{\epsilon'z(1+z)}{3(3-\epsilon')}
\sum_{n=0}^\infty \frac{(\epsilon'+1)_n}{(4-\epsilon')_n}
(-z)^n+O(\epsilon'\,{}^2)\\
&=&1+\frac{z}{3}-\frac{\epsilon'
z}{9}[1+(1+z)\,{}_2F_1(1,1;4;-z)]+O(\epsilon'\,{}^2),
\end{eqnarray*}
where
\[{}_2F_1(1,1;4;-z)=-\frac{3}{z}\bl\{\frac{3}{2}+\frac{1}{z}-\frac{(1+z)^2}{z^2}\,\log\,(1+z)\br\}.\]
This then produces
\[-\frac{1}{\xi} \Im \{z^{\epsilon'-1} {\cal
F}\}=-\frac{1}{\xi}\Im\bl\{z^{\epsilon'-1}\bl(1+\frac{z}{3}+\epsilon'\bl(\frac{5}{6}+\frac{7z}{18}+\frac{1}{3z}\br)-\epsilon'\frac{(1+z)^3}{3z^3}\,\log
(1+z)+O(\epsilon'\,{}^2)\br)\br\}\]
\[=1+\epsilon'\bl(\frac{3}{2}-\frac{1}{2} \log (1+X)-\frac{12-X}{6}
\log\,\frac{4+X}{1+X}+L\br)+O(\epsilon'\,{}^2),\]
where
\[L:=\frac{2}{3\xi}(4-3X)(\phi-\omega)-\frac{4\phi}{3\xi}=\frac{2}{3\xi}\{(2-3X)(\phi-2\omega)-3X\omega\}.\]
Upon noting that
\[\frac{q^{-2\epsilon'}\g(1+\epsilon')}{\pi\epsilon'(1-\epsilon')(2-\epsilon')
(16\pi)^{2-\epsilon'}}=
\frac{1}{512\pi^3
\epsilon'}\bl\{1+\epsilon'\bl(\frac{3}{2}-\gamma+\log\,\frac{16\pi}{q^2}\br)+O(\epsilon'\,{}^2)\br\}\]
and use of the above representation of the quantity $\phi-2\omega$, we finally
obtain after some algebra the behaviour near the upper boundary when $D=1$ given by
\[I_{1,m}(p,q)=\frac{1}{512\pi^3}\bl\{\frac{1}{\epsilon'}+3-\gamma-\log
\frac{\sqrt{4p^2+q^4}}{16\pi}-\frac{2}{3}\bl(3-\frac{p^2}{q^4}\br) \log
\frac{4(p^2+q^4)}{4p^2+q^4}\hspace{1cm}\]
\bee\label{e44}
\hspace{3cm}-\frac{2(q^4-6p^2)}{3pq^2} \arctan
\frac{2p^3/q^2}{q^4+3p^2}-\frac{4p}{q^2} \arctan \frac{p}{q^2} + O(\epsilon')\br\}
\ee
as $\epsilon'\to 0+$.
\vspace{0.2cm}

\noindent{\bf C.2.\ \ The case $D=3$}
\vspace{0.2cm}

First we observe from Fig.~1 that the lower boundary is not approached when $D=3$.
In the neighbourhood of the upper boundary when $D=3$, we put $m=2-2\epsilon'$
and let $\epsilon'\to 0+$. From (\ref{e16}), we have
\[I_{3,m}(p,q)=\frac{q^{-2\epsilon'} \g(1+\epsilon')}{2\pi\epsilon'
(16\pi)^{1-\epsilon'}}\,(1+X)^{-\epsilon'/2}\,{}_2F_1(\fs\epsilon',
1-\fs\epsilon'; \f{3}{2};\zeta),\qquad \zeta:=\frac{X}{1+X}.\]
Proceeding as in the previous section to expand the hypergeometric function, we
obtain
\[
{}_2F_1(\fs\epsilon', 1-\fs\epsilon';
\f{3}{2};\zeta)=1+\f{1}{3}\epsilon'\zeta\,{}_2F_1(1,1;\f{5}{2};\zeta)+O(\epsilon'\,{}^2).
\]
Then, with the evaluation
\[{}_2F_1(1,1;\f{5}{2};\zeta)=\frac{3}{\zeta}\bl(1-\sqrt{\frac{1-\zeta}{\zeta}}
\,\arcsin \sqrt{\zeta}\br)=
\frac{3}{\zeta}\bl(1-\frac{\phi}{\sqrt{X}}\br),\]
where $\phi$ is defined in (\ref{e400}), we obtain
\[
{}_2F_1(\fs\epsilon', 1-\fs\epsilon'; \f{3}{2};\zeta)=
1+\epsilon' \bl(1-\frac{\phi}{\sqrt{X}}\br)+O(\epsilon'\,{}^2).
\]
The remaining algebra is straightforward, leading to the final result describing the
behaviour in the neighbourhood of the upper boundary when $D=3$, given by
\bee\label{e46}
I_{3,m}(p,q)=\frac{1}{32\pi^2}\bl\{\frac{1}{\epsilon'}+1-\gamma-\log\,\frac{\sqrt{4p^2+q^4}}{16\pi}-\frac{q^2}{2p}
\arctan \frac{2p}{q^2}+O(\epsilon')\br\}
\ee
as $\epsilon'\to 0+$. This agrees with \cite[(A.8)]{MC99} and \cite[(5.36),
(5.40)]{RDS11},
while the $O(\epsilon')$ term of the
expansion (\ref{e46}) has been used to produce the second-order contribution
in \cite[(5.40)]{RDS11}.

\vspace{0.6cm}

\begin{center}
{\bf Appendix D: \ Bounds on $F_{nk}$ }
\end{center}
\setcounter{section}{4}
\setcounter{equation}{0}
\renewcommand{\theequation}{\Alph{section}.\arabic{equation}}
We first establish the following lemma:
\begin{lemma}$\!\!\!.$\ Let $k=0, 1, 2, \ldots$ and the parameters $\beta_1>\alpha_1>0$, $\beta_2>\alpha_2>0$. Then the terminating hypergeometric series ${}_3F_2(-k,\alpha_1, \alpha_2;\beta_1, \beta_2;1)$ satisfies
\[0<{}_3F_2\bl(\begin{array}{c}-k, \alpha_1, \alpha_2\\ \beta_1, \beta_2\end{array}\!;1\br)\leq 1.\]
\end{lemma}

\noindent{\it Proof.}\ \ From \cite[(15.6.1)]{DLMF} we have for $\beta_1>\alpha_1>0$
\[{}_2F_1(-k,\alpha_1;\beta_1;t)=\frac{1}{B(\alpha_1, \beta_1-\alpha_1)} \int_0^1u^{\alpha_1-1}(1-u)^{\beta_1-\alpha_1-1} (1-ut)^k du,\]
where $B(a,b)$ is the beta function. Replacement of the beta function by its integral representation then shows that
\[{}_2F_1(-k,\alpha_1;\beta_1;t)=\frac{\int_0^1u^{\alpha_1-1}(1-u)^{\beta_1-\alpha_1-1} (1-ut)^k du}{\int_0^1u^{\alpha_1-1}(1-u)^{\beta_1-\alpha_1-1} du},\]
which, for $t\in [0,1]$, clearly has a value in the interval $(0,1]$. From the Euler integral representation for the ${}_3F_2$ series  \cite[(16.5.2)]{DLMF}, we have for $\beta_2>\alpha_2>0$
\begin{eqnarray*}
F:={}_3F_2\bl(\begin{array}{c}-k, \alpha_1, \alpha_2\\ \beta_1, \beta_2\end{array}\!;1\br)&=&\frac{1}{B(\alpha_2, \beta_2-\alpha_2)}\int_0^1 t^{\alpha_2-1}(1-t)^{\beta_2-\alpha_2-1} \,{}_2F_1(-k,\alpha_1;\beta_1;t)\,dt\\
&=&\frac{\int_0^1 t^{\alpha_2-1}(1-t)^{\beta_2-\alpha_2-1} \,{}_2F_1(-k,\alpha_1;\beta_1;t)\,dt}{\int_0^1 t^{\alpha_2-1}(1-t)^{\beta_2-\alpha_2-1}dt} \in (0,1].
\end{eqnarray*}
Hence, $0<F\leq 1$ provided $\beta_1>\alpha_1>0$, $\beta_2>\alpha_2>0$
and $k=0, 1,2, \ldots\ $.

Identification of the parameters $\alpha_j$, $\beta_j$ ($j=1,2$) with those appearing in $F_{nk}$ in (\ref{c2}),
namely $\alpha_1=\fs(a+n)$, $\alpha_2=\fs(a+n+1)$, $\beta_1=\f{3}{2}+n$ and $\beta_2=b+1+n$, where $n=0, 1, 2, \ldots\,$, we see that $\alpha_1>0$, $\alpha_2>0$ and $\beta_1-\alpha_1=\fs n+\fs(3-a)>0$ for $a=2-\epsilon\in (0,2)$. The remaining condition is
\[\beta_2-\alpha_2=\fs n+b-\fs (a-1)=\fs n+\fs(2-D+\epsilon)=\fs n+\f{1}{4}m>0\]
by (\ref{e110}). The conditions of the lemma are satisfied by $F_{nk}$ in the convergence domain. Hence it follows that
\bee\label{d1}
0<F_{nk}\leq 1
\ee
for non-negative integers $n$ and $k$.

\end{document}